\documentclass[a4paper,11pt]{article}
\pdfoutput=1

\usepackage{jheppub}
\usepackage[T1]{fontenc}

\usepackage[utf8]{inputenc} 
\usepackage{float}
\usepackage{amsmath,amssymb,mathtools} 
\usepackage{graphicx} 
\usepackage{booktabs} 
\usepackage{xcolor} 
\usepackage{hyperref} 
\usepackage{slashed} 
\usepackage{bm}

\title{Asymmetric Reheating of a Dark Sector}

\author[a]{Catarina Cosme,$^1$\note{ORCID: \href{https://orcid.org/0000-0001-5148-8868}{0000-0001-5148-8868}}}
\author[a]{João G. Rosa,$^2$\note{ORCID: \href{https://orcid.org/0000-0001-8257-016X}{0000-0001-8257-016X}}}
\author[a]{and In\^{e}s Sequeira}

\affiliation[a]{Univ Coimbra, Faculdade de Ci\^{e}ncias e Tecnologia da Universidade de Coimbra and CFisUC,
Rua Larga, 3004-516 Coimbra, Portugal}

\emailAdd{ccosme@uc.pt}
\emailAdd{jgrosa@uc.pt}
\emailAdd{sequeirainesrodrigues@gmail.com}

\abstract{
We study asymmetric post-inflationary reheating in a framework with two sequestered MSSM-like sectors (the ``visible'' and ``dark'' sectors) with different supersymmetry-breaking scales, with dark matter corresponding to the lightest dark baryons. In this setup, independent Affleck-Dine dynamics in the two sectors offers a natural explanation for the comparable abundances of visible and dark matter. We relate the resulting temperature hierarchy to the inflaton's partial decay widths in different kinematical regimes, where it may decay into dark sector particles via 2-, 3- or 4-body processes. 
The latter case yields the most asymmetric reheating scenario, and we discuss the implications of this regime for the dark sector's thermal history, namely regarding the corresponding strong and weak interactions. While in this kinematical regime the contribution of light dark sector degrees of freedom to $\Delta N_{\rm eff}$ is negligible, we show that it increases significantly in the regime where 3-body decays are possible, particularly near the threshold for dark Higgs pair-production where $\Delta N_{\rm eff}^{\max} \simeq 0.078$. The current combined BBN+CMB+BAO bound leaves only a narrow allowed interval of the 2-body decay regime immediately above threshold, while next-generation CMB experiments could probe slightly below it.}

\begin{document}

\maketitle

\section{Introduction}
\label{sec:introduction}

The standard cosmological model provides a remarkably successful
description of the evolution of the Universe from an early
radiation-dominated epoch to the present day. Nevertheless, several
fundamental questions remain open, including the origin of the initial
conditions required for this evolution and the nature of dark matter.
Cosmic inflation provides a compelling framework for addressing the former
by introducing a period of accelerated expansion in the early Universe.
The subsequent reheating stage transfers the energy stored in the inflaton
field to the particle degrees of freedom that populate the primordial
plasma \cite{Guth:1980zm,Riotto:2002yw, Allahverdi:2010xz}.

The nature and origin of dark matter remain among the central open problems
of cosmology and particle physics. Dark matter accounts for approximately
$85\%$ of the total matter content of the Universe, with a present-day
abundance about five times larger than that of ordinary baryonic
matter~
\cite{ParticleDataGroup:2024cfk,Planck:2018vyg}. Despite their very different physical properties and,
potentially, different microscopic origins, the baryonic and dark matter
abundances are therefore of the same order, a puzzling observation known as the baryon-dark matter coincidence problem. Asymmetric dark matter
scenarios can relate the number densities of the visible and dark
components, although in many constructions a separate coincidence between
the corresponding particle masses is still required
\cite{Nussinov:1985xr, Gelmini:1986zz, Chivukula:1989qb, Barr:1990ca, Kaplan:1991ah, Hodges:1993yb, Berezhiani:2000gw, Berezhiani:2003xm, Hooper:2004dc, Kitano:2004sv, Cosme:2005sb, Farrar:2005zd, Suematsu:2005kp, Berezhiani:2005ek, Tytgat:2006wy, Falkowski:2006qq, Berezhiani:2008zza, Kaplan:2009ag, An:2009vq, Ciarcelluti:2010zz, March-Russell:2011ang, Cirelli:2011ac, Das:2011gj, Gu:2012fg, Petraki:2013wwa, Zurek:2013wia, Bai:2013xga, Higaki:2013vuv, Gu:2013nya, Newstead:2014jva, GarciaGarcia:2015pnn, Reece:2015lch, Farina:2016ndq, Lonsdale:2018xwd, Ibe:2019yra, Ibe:2019ena, Blinov:2021mdk, Ibe:2021gil, Ritter:2021hgu, Ritter:2022opo, Bodas:2024idn, Chung:2024nnj,Chung:2024ezq, Chang:2025eef, Chung:2025wle}.

In this work, we consider the framework introduced in
Ref.~\cite{Rosa:2022tch}, which provides a possible explanation for the similarity between the
baryonic and dark matter abundances without requiring the particles in the
two sectors to have comparable masses\footnote{Note that other scenarios with this feature have been proposed in the recent literature \cite{Brzeminski:2023wza, Banerjee:2024xhn, Ritter:2024sqv}.}. The model contains two sequestered Minimal Supersymmetric Standard
Model (MSSM)-like sectors with the same gauge structure and common
high-energy physics, but potentially very different supersymmetry (SUSY) breaking
scales.
A baryon-antibaryon asymmetry is generated independently in each sector
through the Affleck-Dine mechanism~
\cite{Affleck:1984fy,Dine:1995uk,Allahverdi:2012ju}. In the visible sector,
this asymmetry accounts for the observed baryonic abundance, while in the
dark sector the corresponding asymmetry determines the dark matter
abundance. The different supersymmetry-breaking scales affect both the
generated asymmetries and the QCD confinement scales, allowing comparable
present-day energy densities even when the characteristic particle masses
in the two sectors are very different. Moreover, a heavy dark sector is naturally more abundant and feebly-interacting than the visible sector, in line with the known properties of dark matter.

A crucial ingredient of this scenario is asymmetric post-inflationary
reheating. Although the inflaton couples to both sectors, the available
decay channels depend on their respective particle spectra. In particular, if, as commonly assumed, the inflaton is a gauge singlet, it cannot couple directly to the chiral fermions in either sector\footnote{ While this is not a strict requirement, gauge interactions may spoil the flatness of the inflaton's potential through radiative corrections, as well as affect the amplitude of the resulting primordial curvature perturbations, which justifies this assumption. We also note that a gauge-singlet inflaton could couple directly to right-handed neutrinos (in both sectors), assuming of course the existence of such states. We will not, however, pursue this possibility in this work.}.
If the
dark sector scalar states are substantially heavier than their visible
counterparts, as a result of the larger dark SUSY breaking scale, inflaton decays into the dark sector may be kinematically
suppressed, leading to a smaller radiation energy density and a lower
temperature after reheating. We characterize this asymmetry through the
ratio $\epsilon=\Gamma'/\Gamma$ of the partial decay widths of the inflaton into
the dark and visible sectors, respectively.

The efficiency of dark sector reheating depends in particular on the
hierarchy between the inflaton mass, $m_\phi$, and the dark Higgs mass,
$M_{H'}$, which we take to be the lightest dark scalar (given the typical loop-factor suppression of its mass relative to the SUSY breaking scale). Three distinct kinematic regimes arise. For
$m_\phi>2M_{H'}$, the inflaton can decay directly into two on-shell dark
Higgs bosons, and the dominant dark sector decay is a 2-body process. At
the opposite end, for $m_\phi<M_{H'}$, neither dark Higgs can be produced
on shell and the inflaton decay proceeds through two virtual scalar
mediators into two dark fermion-antifermion pairs, resulting in a strongly
suppressed 4-body rate.

The intermediate regime, $M'_H<m_\phi<2M'_H$, lies between these
two limiting cases. In this mass range, the single-resonant
3-body topology becomes kinematically accessible for
$m_\phi\geq M'_H+2m'_f$, allowing one dark Higgs boson to be
produced on-shell while the second remains virtual and subsequently
decays (dominantly) into a dark fermion-antifermion pair. The resulting 3-body channel therefore
interpolates between the fully on-shell 2-body process and the 4-body
decay mediated by two virtual scalars. A consistent treatment of this regime requires retaining the
final-state fermion mass since
$m_{f'}=y'M_{H'}$, so that the effective Yukawa coupling $y'$ affects both the matrix element and
the available phase space. Moreover, as $m_\phi$ approaches $2M_{H'}$ from below, the
virtual dark Higgs approaches its mass shell and its finite decay width
must be taken into account.

We study the dynamics of asymmetric reheating by solving the Boltzmann
equations governing the inflaton and radiation energy densities in both
sectors. We derive analytical expressions for their evolution and for the
resulting temperature hierarchy, and determine the corresponding reheating
and maximum temperatures. We then analyze the subsequent dark sector thermal history in the
strongly hierarchical 4-body regime, including the possibility of
a dark QCD crossover and the thermalization of dark weak
interactions. Finally, we investigate
the contribution of the dark radiation bath to the effective number of
relativistic neutrino species, $\Delta N_{\rm eff}$.

We find that the different kinematic regions lead to qualitatively
different cosmological outcomes. When both dark Higgs bosons are
off shell, the decay into the dark sector is strongly suppressed,
yielding an extremely cold dark sector and a negligible contribution
to $\Delta N_{\rm eff}$. Once the single-resonant 3-body channel
becomes dominant, the dark radiation contribution grows rapidly as
the dark Higgs pair-production threshold is approached. In the immediate vicinity of $m_\phi=2M_{H'}$, we use an approximate finite-width treatment of the four-fermion final state to interpolate smoothly between the 3- and 2-body descriptions. For representative benchmark parameters, this gives $\Delta N_{\rm eff}^{\max}\simeq 0.078$ at threshold.
For this benchmark, the current combined BBN+CMB+BAO bound excludes
the 2-body regime except for a narrow allowed interval immediately
above the threshold. A future CMB experiment reaching a sensitivity
of order $\Delta N_{\rm eff}\simeq0.06$ would also be sensitive to a
small region below the threshold.
This transition therefore provides
an observationally interesting probe of asymmetric dark sector
reheating.

This paper is organised as follows. In
Section~\ref{sec:model}, we introduce the two-sector framework and
summarise the generation of the visible and dark asymmetries through
the Affleck-Dine mechanism. In Section~\ref{sec:reheating}, we study
the dynamics of asymmetric reheating and derive the evolution of the
energy densities and temperatures in both sectors. In
Section~\ref{sec:decays}, we analyse the relevant inflaton decay
channels, including a detailed treatment of the 3-body decay in
the intermediate mass regime. In Section~\ref{sec:cosmology}, we analyse the dark sector thermal history in the
strongly hierarchical 4-body regime, including dark QCD and dark
weak interactions, while Section~\ref{sec:neff} discusses the constraints from
$\Delta N_{\rm eff}$ and compares the different inflaton decay
regimes. We present our conclusions in
Section~\ref{sec:conclusions}. The detailed analytical solutions and
decay-width calculations are included in the appendices.


\section{The two-sector framework}
\label{sec:model}

We consider the framework proposed in Ref.~\cite{Rosa:2022tch},
which contains two sequestered sectors: the visible sector, comprising
ordinary baryonic matter, and a dark sector containing the dark matter
component. The two sectors interact only very weakly, consistent
with the absence of sizeable non-gravitational interactions between
visible and dark matter. Each sector contains the complete particle
content of the MSSM and has the same gauge structure, while the
corresponding supersymmetry-breaking scales are allowed to differ.

This setup provides a possible explanation for the similarity between
the present-day baryonic and dark matter abundances. The
Affleck-Dine mechanism generates a matter-antimatter asymmetry
independently in each sector, while the different SUSY-breaking
scales determine both the magnitude of these asymmetries and the
characteristic baryon masses. A larger dark supersymmetry-breaking scale may therefore be accompanied by a smaller
dark baryon asymmetry, with the two effects combining to yield
comparable visible- and dark sector energy densities
\cite{Rosa:2022tch}.


\subsection{Visible and dark sectors}

The assumption that the two sectors share the same gauge structure and
high-energy physics can be motivated by heterotic string constructions,
such as those based on an $E_8 \times E_8$ symmetry ~\cite{Rosa:2022tch}. Such theories
naturally contain two copies of
$SU(3)\times SU(2)\times U(1)$ above the unification scale. After the
larger symmetry is spontaneously broken, the two copies evolve as
independent MSSM-like sectors. 

We therefore assume a common unification scale, $M_{\rm GUT}$, at
which the visible and dark gauge couplings satisfy the same boundary
conditions,
\begin{equation}
    \alpha_i^{-1}(M_{\rm GUT})
    =
    \alpha_i^{\prime\,-1}(M_{\rm GUT})
    \simeq 25.6,
    \qquad
    i=1,2,3,
\end{equation}
with
\begin{equation}
    M_{\rm GUT}\sim 2\times 10^{16}\ {\rm GeV},
\end{equation}
as assumed in Ref.~\cite{Rosa:2022tch}. Here and henceforth, we use primed (unprimed) quantities when referring to the dark (visible) sectors.

Although the two sectors are identical at high energies, they may
develop different low-energy spectra if supersymmetry breaking is
communicated to them with different strengths.

We assume that supersymmetry is broken in a hidden sector by a field
$\chi$ acquiring a non-vanishing $F$-term, $F_\chi$. The breaking is communicated to the visible and dark sectors through
messenger fields. 
If the messenger couples differently to the two sectors, their
supersymmetry-breaking scales are given parametrically by
\begin{equation}
    M_{\rm SUSY}
    \sim
    \lambda\,\frac{F_\chi}{M},
    \qquad
    M'_{\rm SUSY}
    \sim
    \lambda'\,\frac{F_\chi}{M},
\end{equation}
where $\lambda$ and $\lambda'$ denote the corresponding messenger
couplings \cite{Martin:1997ns,Rosa:2022tch}. The scalar masses in each sector are then expected to be of
the order of the respective supersymmetry-breaking scale. For the
visible sector, we take\footnote{Note that this is simply a convenient benchmark value, since the scenario only depends on the ratio $M'_{\rm SUSY}/M_{\rm SUSY}$. In fact, gauge coupling unification may be possible for $M_{\rm SUSY}=\mathcal{O}(1-100)$ TeV \cite{Bhattiprolu:2023lfh}}
\begin{equation}
    M_{\rm SUSY}\sim \mathcal{O}(1)\ {\rm TeV},
\end{equation}
whereas the dark supersymmetry-breaking scale may be considerably
larger.

In the visible sector, the present matter density is dominated by
protons and neutrons, which are the lightest stable baryons. By
analogy, we assume that the dark matter density is dominated by the
lightest stable dark baryons.


\subsection{Dark baryon abundance}
\label{subsec:dark-abundance}

Since both sectors contain the MSSM field content, the Affleck-Dine
mechanism can operate independently in each of them~\cite{Affleck:1984fy,Dine:1995uk,Allahverdi:2012ju}. In each
sector, a complex scalar field $\Phi_{\rm AD}$ parametrising an MSSM flat
direction \cite{Gherghetta:1995dv} acquires a large field value during inflation due to
Hubble-induced supersymmetry-breaking effects. As the Hubble rate decreases
to $H\sim m_{\rm AD}$, the field begins to oscillate and the relative phase
between the Hubble-induced and soft $A$-terms generates a rotational motion
of the condensate, producing a non-vanishing baryon asymmetry. The
subsequent decay of the condensate transfers this asymmetry to the quark
sector.

For a flat direction lifted by a non-renormalisable superpotential operator
of order $n$, the resulting baryon-to-entropy ratio in either sector is
approximately~\cite{Dine:1995uk,Rosa:2022tch}
\begin{equation}
\eta_B \equiv \frac{n_B}{s}
\simeq
\beta\,\frac{n-2}{6(n-3)}
\left(
\frac{M^{n-3}}
{\lambda M_{\rm Pl}^{n-2}}
\right)^{\delta_n}
T_{\rm RH} M_{\rm SUSY}^{\delta_n-1},
\qquad
\delta_n \equiv \frac{2}{n-2}.
\label{etaB}
\end{equation}
where $\beta$ denotes the baryon charge of the flat direction, $\lambda$ is
the coupling of the non-renormalisable operator, $M$ is the scale suppressing
it, and $T_{\rm RH}$ is the reheating temperature. Here, $M_{\rm Pl}\simeq 2.4\times10^{18}\,\mathrm{GeV}$ denotes the
reduced Planck mass, while $s$ denotes the
total entropy density after reheating, which is dominated by the
visible-sector radiation bath in the asymmetric-reheating scenario
considered below.

In the MSSM with unbroken $R$-parity, the flat directions relevant for
baryon-number generation are lifted by operators with $n=4$ or $n=6$.
Since $n=6$ flat directions typically acquire larger field values, we focus
on this case~\cite{Dine:1995uk,Rosa:2022tch}. For instance, taking $M\sim M_{\rm Pl}$ and $\lambda\sim 1$, one obtains the observed visible-sector baryon asymmetry,
$\eta_B\sim10^{-10}$, for a reheating temperature of the order of the electroweak scale, $T_{\rm RH}\sim100~{\rm GeV}$.
%
%

Assuming the same high-energy parameters in the two sectors, the resulting
baryon asymmetries are, hence, related by~\cite{Rosa:2022tch}
\begin{equation}
\frac{\eta'_B}{\eta_B}
\sim
\left(
\frac{M_{\rm SUSY}}{M'_{\rm SUSY}}
\right)^{1/2}.
\label{etaratio}
\end{equation}
Thus, for $M'_{\rm SUSY}>M_{\rm SUSY}$, the dark sector baryon asymmetry
is smaller than its visible-sector counterpart. The dependence on the
hierarchy between the two supersymmetry-breaking scales is, however,
relatively mild.

The different supersymmetry-breaking scales also modify the baryon
masses through their effect on the QCD confinement scale.  Under the
assumptions described in Ref.~\cite{Rosa:2022tch}, the dark QCD scale can be estimated as
\begin{equation}
\Lambda'_{\rm QCD}
=
{M'_{\rm SUSY}}^{1-\kappa}
{M_{\rm GUT}}^\kappa
\exp\!\left[
-\frac{2\pi}{b_5}
\left(\alpha_{\rm GUT}^{-1}-1\right)
\right]
\left(\frac{m'_t}{M'_{\rm SUSY}}\right)^{1-b_6/b_5}.
\end{equation}
where $b_6=7$ and $b_5=23/3$ are the one-loop QCD
beta-function coefficients with six and five active quark flavours,
respectively, while $b_{\rm MSSM}=3$ is the corresponding MSSM
coefficient above the supersymmetry-breaking scale. We define
\begin{equation}
    \kappa \equiv \frac{b_{\rm MSSM}}{b_5}
    = \frac{9}{23}.
\end{equation}

For sufficiently light quarks, the baryon masses are approximately
proportional to the corresponding confinement scale \cite{Rosa:2022tch},
\begin{equation}
    m_B\sim 5\Lambda_{\rm QCD},
    \qquad
    m'_B\sim 5\Lambda'_{\rm QCD}.
\end{equation}
It is natural to assume an $\mathcal{O}(1)$ dark top Yukawa coupling,
as in the MSSM. If the dark electroweak scale is set by
$M'_{\rm SUSY}$, we then expect
\begin{equation}
    m'_t \sim \mathcal{O}(1)\Lambda'_{\rm EW},
    \qquad
    \frac{m_t}{M_{\rm SUSY}}
    \sim
    \frac{m'_t}{M'_{\rm SUSY}} .
    \label{eq:darkquarkmass}
\end{equation}
The ratio between the dark and visible baryonic energy densities then
becomes \cite{Rosa:2022tch}
\begin{equation}
   \frac{\rho'_B}{\rho_B}
=
\frac{m'_B\eta'_B}{m_B\eta_B}
=
\frac{\Lambda'_{\rm QCD}\eta'_B}
{\Lambda_{\rm QCD}\eta_B}
=
\left(\frac{M'_{\rm SUSY}}{M_{\rm SUSY}}\right)^{\delta_n-\kappa}
\times \mathcal{O}(1).
\end{equation}
For $n=6$, the exponent is therefore small,
\begin{equation}
    \delta_6-\kappa\simeq 0.1,
\end{equation}
so that the final abundance ratio depends only weakly on the hierarchy
between the two supersymmetry-breaking scales. Consequently, even if
the dark baryons are substantially heavier than their visible-sector counterparts, the two energy densities remain of the same order. In particular, $\rho'_B/\rho_B$ only reaches values of order ten for a very large hierarchy,
\begin{equation}
    \frac{M'_{\rm SUSY}}{M_{\rm SUSY}}
    {\lesssim} 10^{12}.
\end{equation}
The framework can therefore account for comparable baryonic and dark
matter abundances without requiring comparable particle masses \cite{Rosa:2022tch}.

As in Ref.~\cite{Rosa:2022tch}, we assume that the lightest dark baryon is
sufficiently stable to constitute the dominant dark-matter component,
while any relic abundance associated with supersymmetric particles is
negligible. In particular, an exactly conserved $R$ parity would imply
a stable lightest supersymmetric particle, which could provide an
additional dark-matter component. This may be avoided, for example, in
models with broken $R$ parity in which proton and dark-baryon stability
are instead ensured by an additional symmetry. We do not specify this
ultraviolet completion here and assume that the baryonic component
dominates the present dark-matter abundance.

The hierarchy between the two supersymmetry-breaking scales also has
important consequences for reheating. We assume that the inflaton is a gauge singlet and may decay
into the degrees of freedom of both sectors. However, if the dark scalar particles are much heavier than
their visible-sector counterparts, inflaton decays into dark scalars
may be kinematically forbidden or strongly suppressed \cite{Rosa:2022tch}. Decays into
chiral dark fermions must then proceed through virtual scalar mediators.

The inflaton consequently transfers energy less efficiently to the
dark sector than to the visible sector, where decays into the (much) lighter
on-shell scalars remain allowed (assuming the inflaton's mass is well above the TeV scale). This naturally produces different
radiation densities and temperatures in the two sectors. The dynamics
of this asymmetric reheating process will be analysed in detail in
Section~\ref{sec:reheating}.

\section{Asymmetric reheating}
\label{sec:reheating}

We now consider the post-inflationary evolution of the two sectors
introduced in Section~\ref{sec:model}. Since the inflaton decays into
the visible and dark sectors with different partial decay widths, its
energy is transferred unevenly between the two radiation baths. This
naturally leads to different energy densities and temperatures after
inflation.

At the end of inflation, the energy density of the Universe is assumed
to be dominated by the inflaton field. We assume that, around the
minimum of its potential, the inflaton potential is approximately
quadratic. The coherently oscillating inflaton condensate then has an
average equation of state $w_\phi\simeq0$ and behaves effectively as
non-relativistic matter while gradually decaying into particles
belonging to the two sectors. We assume that the energy deposited by
inflaton decays is rapidly transferred to relativistic degrees of
freedom in each sector, so that the corresponding energy densities can
be treated as radiation~\cite{Turner:1983he,Allahverdi:2010xz}.

\subsection{Boltzmann equations}
\label{subsec:boltzmann}

The evolution of the inflaton and radiation energy densities is
described by the usual perturbative-reheating Boltzmann equations \cite{Allahverdi:2010xz}:
\begin{align}
    \dot{\rho}_{\phi}+3H\rho_{\phi}
    &=
    -\Gamma_{\phi}\rho_{\phi},
    \label{eq:boltzmann-inflaton}
    \\
    \dot{\rho}+4H\rho
    &=
    \Gamma\rho_{\phi},
    \label{eq:boltzmann-visible}
    \\
    \dot{\rho}'+4H\rho'
    &=
    \Gamma'\rho_{\phi},
    \label{eq:boltzmann-dark}
\end{align}
where $\rho_{\phi}$ denotes the inflaton energy density, while $\rho$
and $\rho'$ are the radiation energy densities in the visible and dark
sectors, respectively. The corresponding partial decay widths are
denoted by $\Gamma$ and $\Gamma'$, and the total inflaton decay width is
\begin{equation}
    \Gamma_{\phi}=\Gamma+\Gamma'.
    \label{eq:total-width}
\end{equation}
Here we assume that any massive particles produced directly in inflaton
decays either decay or annihilate sufficiently rapidly into lighter
relativistic degrees of freedom. We further assume that the relativistic
degrees of freedom constituting each radiation bath establish internal
thermal equilibrium. This does not require every particle species in the
dark sector to thermalise with this bath; the efficiency of dark weak
interactions is considered separately in Sec.~\ref{subsec:dark-weak-interactions}. Under these
assumptions, both $\rho$ and $\rho'$ obey radiation-like equations of
state during reheating. This relation ensures conservation of the total
energy density,
\begin{equation}
    \dot{\rho}_{\rm tot}
    +
    3H
    \left(
        \rho_{\phi}
        +\frac{4}{3}\rho
        +\frac{4}{3}\rho'
    \right)
    =
    0,
    \label{eq:total-energy-conservation}
\end{equation}
with
\begin{equation}
    \rho_{\rm tot}
    =
    \rho_{\phi}+\rho+\rho'.
\end{equation}
The different coefficients reflect the equations of state of the
three components: the oscillating inflaton behaves as matter, whereas the particles produced in its decay behave as radiation.

The expansion rate is determined by
\begin{equation}
    H
    =
    \left(
        \frac{\rho_{\phi}+\rho+\rho'}
             {3M_{\rm Pl}^{2}}
    \right)^{1/2}.
    \label{eq:hubble-reheating}
\end{equation}
We take the beginning of reheating to coincide with the end of
inflation, at $t=t_{\rm e}$. Since any pre-existing particle
abundances are strongly diluted during inflation, the initial
conditions are
\begin{equation}
    \rho_{\phi}(t_{\rm e})=V(\phi_{\rm e}),
    \qquad
    \rho(t_{\rm e})=\rho'(t_{\rm e})=0,
    \label{eq:reheating-initial-conditions}
\end{equation}
where the inflation scale is taken, for concreteness, to be $V(\phi_{\rm e})^{1/4}\sim 10^{16}\ {\rm GeV}$ as characteristic of most single-field slow-roll inflation scenarios.

\subsection{Analytical evolution}
\label{subsec:analytical-solutions}

The reheating dynamics may be divided into two approximate regimes \cite{Turner:1983he,Allahverdi:2010xz}.
At early times, the inflaton still dominates the total energy density,
and the Universe behaves as matter-dominated,
\begin{equation}
    a(t)\propto t^{2/3}.
\end{equation}
As the inflaton decays, the radiation produced in the two sectors
gradually becomes dominant. The Universe then enters a
radiation-dominated epoch,
\begin{equation}
    a(t)\propto t^{1/2}.
\end{equation}
The transition occurs around the inflaton lifetime,
\begin{equation}
    \tau_{\phi}\equiv\Gamma_{\phi}^{-1}.
    \label{eq:inflaton-lifetime}
\end{equation}

During inflaton domination, the inflaton energy density evolves approximately as

\begin{equation}
    \rho_{\phi}(t)
    \simeq
    \frac{4M_{\rm Pl}^{2}}{3t^{2}}
    \exp\left[
        -\Gamma_{\phi}
        \left(t-t_{\rm e}\right)
    \right].
    \label{eq:rho-phi-matter}
\end{equation}
For the leading early-time radiation solution, valid for
$\Gamma_\phi t\ll1$, we neglect the exponential depletion of the
inflaton in the source term and use
$\rho_\phi\simeq4M_{\rm Pl}^2/(3t^2)$. The radiation densities sourced
by inflaton decay are then
\begin{align}
\rho(t)
&\simeq
\frac{4 M_{\rm Pl}^2}{5t}\,
\Gamma
\left[
1-\left(\frac{t_e}{t}\right)^{5/3}
\right],
\label{eq:rho-visible-matter}
\\
\rho'(t)
&\simeq
\frac{4 M_{\rm Pl}^2}{5t}\,
\Gamma'
\left[
1-\left(\frac{t_e}{t}\right)^{5/3}
\right]. \label{eq:rho-dark-matter}
\end{align}
Eqs.~(\ref{eq:rho-visible-matter}) and~(\ref{eq:rho-dark-matter}) correspond to the leading early-time
approximation, valid for $\Gamma_\phi t\ll 1$; close to
$t\sim\tau_\phi$, the full numerical evolution must be used. After their initial growth, and while the inflaton still dominates,
the radiation densities decrease approximately as $t^{-1}$, rather than
as $a^{-4}$, because they are continuously replenished by inflaton decays.

Once radiation dominates, the inflaton contribution becomes
exponentially suppressed, while both radiation densities redshift as
\begin{equation}
    \rho,\rho'\propto t^{-2}\propto a^{-4}.
\end{equation}
The complete analytical solutions in the two regimes are presented in
Appendix~\ref{app:boltzmann}.

Within the reheating description above, the structure of
Eqs. (\ref{eq:boltzmann-visible}) and (\ref{eq:boltzmann-dark}), together with the
initial conditions $\rho(t_e)=\rho'(t_e)=0$, implies that the ratio of
the radiation densities remains fixed by the ratio of the corresponding
partial decay widths,
\begin{equation}
    \frac{\rho'}{\rho}
    =
    \frac{\Gamma'}{\Gamma}.
    \label{eq:energy-density-ratio}
\end{equation}
It is therefore convenient to define the reheating asymmetry parameter
as
\begin{equation}
    \epsilon
    \equiv
    \frac{\Gamma'}{\Gamma},
    \label{eq:epsilon}
\end{equation}
so that
\begin{equation}
    \rho'=\epsilon\,\rho.
    \label{eq:rho-prime-epsilon}
\end{equation}
The entire asymmetry between the two radiation baths is thus encoded in
the relative efficiency with which the inflaton decays into each
sector.

\begin{figure}[H] 
\centering 
\includegraphics[pagebox=cropbox,width=0.75\textwidth] {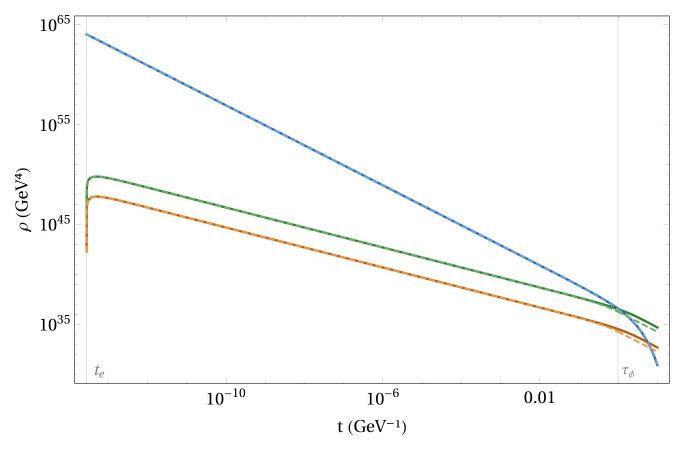} 
\caption{ Evolution of the inflaton energy density (blue), $\rho_\phi$, and of the visible (green) and dark sector (orange) radiation energy densities, $\rho$ and $\rho'$, respectively, for $\epsilon=10^{-2}$ and $\Gamma=1~{\rm GeV}$. Solid curves represent the numerical solutions of the Boltzmann equations, while dashed curves correspond to the analytical approximations. } 
\label{fig:energy-densities} 
\end{figure}

The parameter values used in Figures~\ref{fig:energy-densities} and~\ref{fig:sector-temperatures} serve only to illustrate
the generic reheating dynamics; in particular, $\Gamma=1~\mathrm{GeV}$
is not used in the subsequent phenomenological analysis. The numerical evolution of the three energy densities is shown in Figure~\ref{fig:energy-densities}, together with the analytical approximations derived above. At early times, the inflaton dominates the total energy density, while its decays gradually populate the visible and dark radiation baths. Around the inflaton lifetime, the radiation components become dominant and subsequently redshift as $a^{-4}$. The analytical solutions provide a good description of the numerical evolution in both the inflaton and radiation-dominated regimes. 
\\

\subsection{Temperature asymmetry}
\label{subsec:temperature-asymmetry}

The radiation energy densities are related to the temperatures of the
two sectors through
\begin{equation}
    \rho
    =
    \frac{\pi^{2}}{30}
    g_{*}(T)T^{4},
    \qquad
    \rho'
    =
    \frac{\pi^{2}}{30}
    g'_{*}(T')T'^{4},
    \label{eq:radiation-densities}
\end{equation}
where $g_{*}$ and $g'_{*}$ denote the corresponding effective numbers
of relativistic degrees of freedom. The quantities $T$ and $T'$ denote
the physical temperatures of the internally thermalised visible- and
dark sector radiation baths, respectively. Combining these expressions
with Eq.~\eqref{eq:rho-prime-epsilon} gives
\begin{equation}
    \frac{T'}{T}
    =
    \epsilon^{1/4}
    \left[
        \frac{g_{*}(T)}
             {g'_{*}(T')}
    \right]^{1/4}.
    \label{eq:temperature-ratio}
\end{equation}
The temperature hierarchy is therefore controlled primarily by the
fourth root of the decay-width ratio. In particular, the dark sector
reheating temperature scales approximately as
\begin{equation}
    T'_{\rm RH}
    \sim
    \epsilon^{1/4}T_{\rm RH},
    \label{eq:dark-reheating-temperature-scaling}
\end{equation}
where $T_{\rm RH}$ denotes the visible-sector reheating temperature,
roughly defined at the epoch when $H \simeq \Gamma_\phi$.

Additional relativistic degrees of freedom in the dark sector
contribute to the total radiation density of the Universe. Such a
contribution is naturally suppressed if the dark sector is colder than
the visible one. We therefore focus on
\begin{equation}
    \epsilon<1,
\end{equation}
which implies $T'<T$ when the effective number of relativistic degrees of freedom is comparable.

\begin{figure}[H]
    \centering
    \includegraphics[width=0.75\textwidth]
    {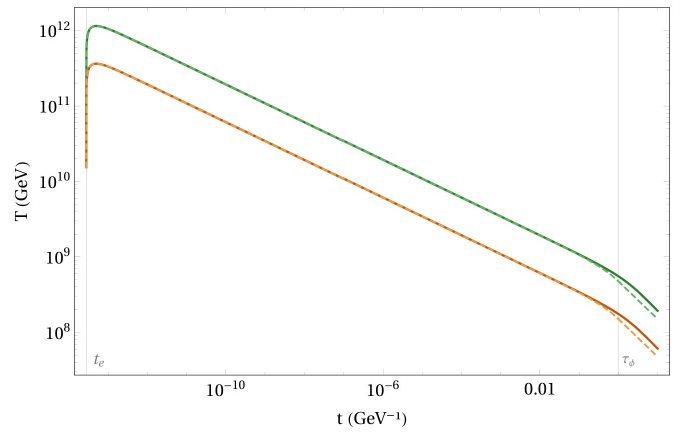}
    \caption{
    Evolution of the visible-sector temperature, $T$ (green), and of the
    dark sector temperature, $T'$ (orange), for $\epsilon=10^{-2}$,
    $\Gamma=1~{\rm GeV}$ and $g_*=g_*'=100$. Solid curves represent
    the numerical solutions, while dashed curves correspond to the
    analytical approximations.
    }
    \label{fig:sector-temperatures}
\end{figure}

The corresponding temperature evolution is shown in
Figure~\ref{fig:sector-temperatures}. Both sectors initially heat up
as the inflaton transfers energy to the radiation baths, reaching a
maximum temperature shortly after the end of inflation. Their
temperatures subsequently decrease as the Universe expands. For the
asymmetric case shown, the dark sector remains colder than the visible
sector throughout the reheating process.
\\

The temperatures do not decrease monotonically from the beginning of
reheating. As the inflaton starts to decay, the radiation densities
initially increase, and each sector reaches a maximum temperature before
the subsequent dilution caused by expansion becomes dominant \cite{Giudice:2000ex, Allahverdi:2010xz}. In
the dark sector, this maximum is determined by
\begin{equation}
    \dot{\rho}'=0.
\end{equation}
Since it is reached during inflaton domination, one obtains
\begin{equation}
    {T'_{\rm max}}^{4}
    =
    \frac{30}
         {\pi^{2}g'_{*}(T'_{\rm max})}
    \frac{3M_{\rm Pl}^{2}}{4}
    \Gamma'H.
    \label{eq:dark-maximum-temperature}
\end{equation}
Using the leading early-time solution above, the maximum occurs at
\begin{equation}
    t_{\rm max}
    =
    \left(\frac{8}{3}\right)^{3/5} t_e .
    \label{eq:tmax}
\end{equation}
During the inflaton-dominated epoch, $H=2/(3t)$, so that
\begin{equation}
    H_{\rm max}
    =
    \left(\frac{3}{8}\right)^{3/5} H_e
    =
    \left(\frac{3}{8}\right)^{3/5}
    \frac{V(\phi_e)^{1/2}}{\sqrt{3}M_{\rm Pl}} .
    \label{eq:Hmax}
\end{equation}
Assuming that $g'_\ast$ does not vary appreciably around the maximum,
the maximum dark sector temperature is therefore
\begin{align}
T'_{\rm max}
&\simeq
\left[
\frac{30\sqrt{3}}
{4\pi^2 g'_\ast(T'_{\rm max})}
\left(\frac{3}{8}\right)^{3/5}
\right]^{1/4}
(M_{\rm Pl}\Gamma')^{1/4}
V(\phi_e)^{1/8}
\nonumber\\
&\simeq
1.15\times10^{12}\ {\rm GeV}
\left[
\frac{100}{g'_\ast(T'_{\rm max})}
\right]^{1/4}
\left(
\frac{\Gamma'}{1\ {\rm GeV}}
\right)^{1/4}
\left[
\frac{V(\phi_e)^{1/4}}
{10^{16}\ {\rm GeV}}
\right]^{1/2}.
\label{eq:Tmaxdark}
\end{align}
The values of $\Gamma$ and $\Gamma'$, and hence of $\epsilon$,
$T'_{\rm RH}$, and $T'_{\rm max}$, depend on the inflaton decay
channels available in each sector. We analyse these channels in the
following section.


\section{Inflaton decay channels}
\label{sec:decays}

The reheating asymmetry introduced in
Section~\ref{sec:reheating} is determined by the partial inflaton decay widths into the visible and dark sectors. These widths depend on the particle spectra of the two sectors and, in particular, on the
hierarchy between the inflaton mass and the corresponding scalar
masses.

We assume that the inflaton is a gauge singlet. This avoids potentially
large corrections from Standard Model gauge interactions to the
inflaton potential and to the amplitude of the primordial curvature
perturbations. At the same time, it prevents direct inflaton decays
into chiral fermions through the usual Yukawa interactions. The
dominant decay channels must therefore involve scalar fields.

Additional gauge-singlet fermions, such as the right-handed neutrinos
considered in Sec.~\ref{subsec:dark-neutrinos}, could in principle couple directly to the inflaton.
We assume that such couplings are absent or negligible, or that the
corresponding decays are kinematically forbidden, so that the scalar
channels considered below dominate reheating.

For concreteness, we adopt an effective single-scalar description of each Higgs sector. The fields $H$ and $H'$ denote representative neutral, CP-even real scalar mass eigenstates of the visible and dark Higgs sectors, with masses $M_H$ and $M_{H'}$, respectively. We assume that these states provide the dominant scalar channels for inflaton decay.
The interactions relevant
for the decay analysis are parametrised as
\begin{equation}
    \mathcal{L}_{\rm int}
    \supset
    -\frac{1}{2}g v_\phi\,\phi H^2
    -\frac{1}{2}g' v_\phi\,\phi H'^2
    -y'H'\overline{f'}f' ,
  \label{eq:inflaton-higgs-interaction}
\end{equation}
where $g$ and $g'$ are dimensionless inflaton couplings, $v_\phi$
denotes the inflaton vacuum expectation value, and $y'$ is the
effective Yukawa coupling of $H'$ to a dark Dirac fermion $f'$.
The detailed Higgs spectra and mixing angles are not modelled
explicitly; their effects are encoded in the effective masses and
couplings appearing above. The decay channel available in each sector
is then determined by the relation between $m_\phi$ and the
corresponding scalar mass.

The three possible kinematic regimes may be summarised as
\begin{equation}
    \phi\rightarrow
    \begin{cases}
        H'H',
        &
        m_{\phi}>2M_{H'},
        \\[2mm]
        H'H'^{*}\rightarrow H'f'\bar f',
        &
        M_{H'}<m_{\phi}<2M_{H'},
        \\[2mm]
        H'^{*}H'^{*}\rightarrow
        f'\bar f'f'\bar f',
        &
        m_{\phi}<M_{H'}.
    \end{cases}
    \label{eq:dark-decay-regimes}
\end{equation}
The expressions above summarise the characteristic decay topology in
each mass range. In the intermediate region
$M'_H<m_\phi<2M'_H$, however, the number of off-shell propagators also
depends on the dark fermion mass, or equivalently on the Yukawa
coupling $y'$. The single-resonant 3-body topology is accessible
only when $m_\phi\geq M'_H+2m'_f$; otherwise the four-fermion final
state remains allowed but both dark Higgs propagators are necessarily
off shell. We discuss this condition explicitly below.

\begin{figure}[!htbp]
    \centering
\includegraphics[width=0.75\textwidth,
    trim=0 270bp 0 300bp,
    clip] {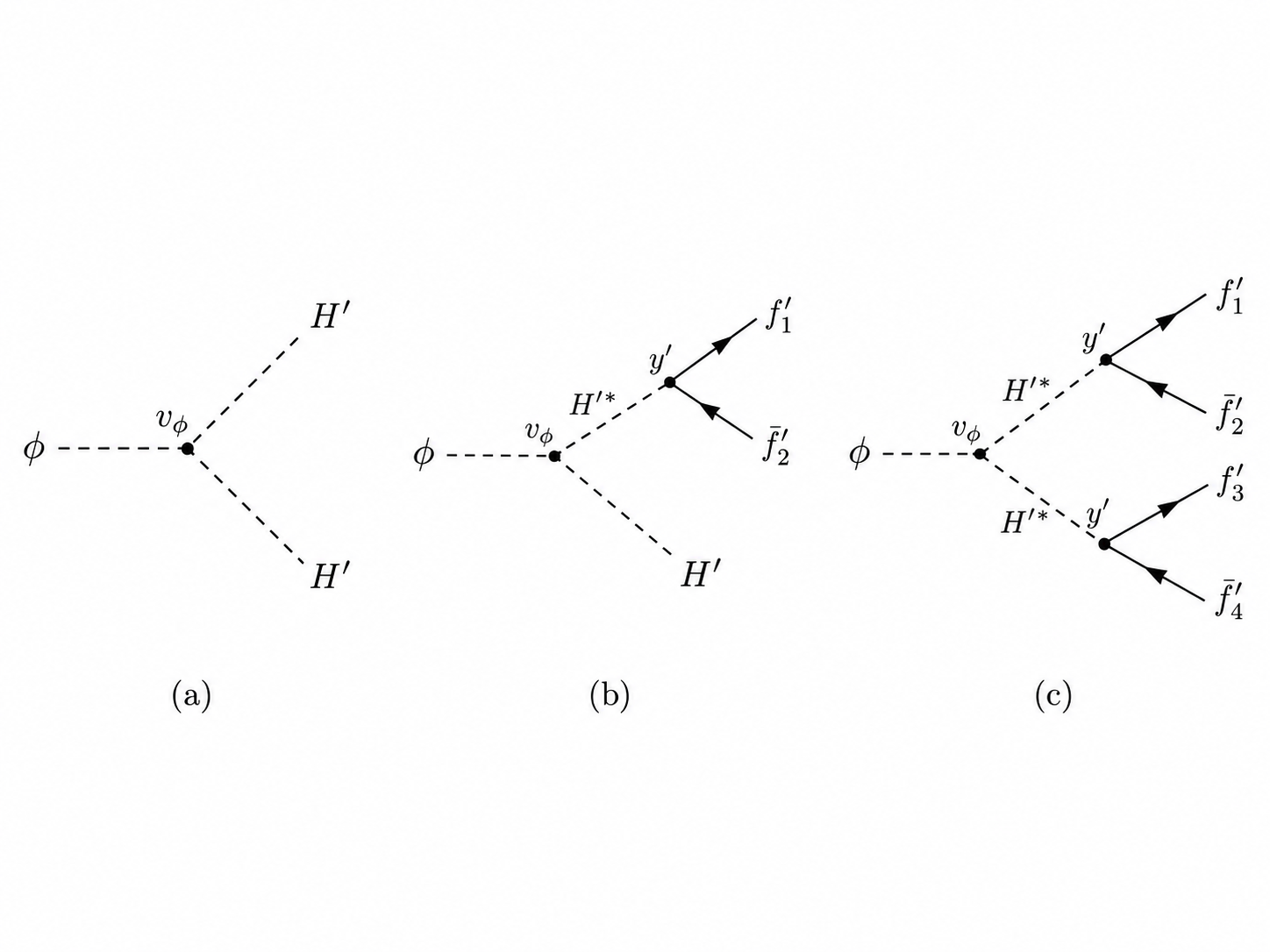}
    \caption{Inflaton decay channels into the dark sector in the three kinematic regimes. Panel (a) shows the 2-body decay into two on-shell dark Higgs bosons. Panel (b) shows the 3-body decay involving one on-shell and one off-shell dark Higgs boson. Panel (c) shows the 4-body decay mediated by two off-shell dark Higgs bosons.}
    \label{fig:inflaton-decay-channels}
\end{figure}

The characteristic decay topologies are illustrated in Figure~\ref{fig:inflaton-decay-channels}.
The mass hierarchy determines which configurations can become
kinematically accessible, while in the intermediate region the actual
number of off-shell dark Higgs propagators also depends on the
dark fermion mass.


\subsection{2-body decay}
\label{subsec:2-body}

We begin with the simplest situation, in which both scalar particles
can be produced on shell. For a decay of the form
\begin{equation}
    \phi\rightarrow HH,
\end{equation}
the kinematic condition is
\begin{equation}
    m_{\phi}>2M_H.
\end{equation}
The corresponding decay width is
\begin{equation}
    \Gamma_{\phi\rightarrow HH}
    =
    \frac{g^{2}v_{\phi}^{2}}
         {32\pi m_{\phi}}
    \left(
        1-\frac{4M_H^{2}}{m_{\phi}^{2}}
    \right)^{1/2}.
    \label{eq:2-body-width-visible}
\end{equation}
For $M_H\ll m_{\phi}$, this reduces to
\begin{equation}
    \Gamma_{\phi\rightarrow HH}
    \simeq
    \frac{g^{2}v_{\phi}^{2}}
         {32\pi m_{\phi}}.
    \label{eq:2-body-width-visible-limit}
\end{equation}
The complete derivation is presented in
Appendix~\ref{app:2-body}.

In the visible sector, we assume the hierarchy
\begin{equation}
    M_{\rm SUSY}\ll m_{\phi},
\end{equation}
so that the decay into a pair of visible Higgs bosons is kinematically
allowed and provides the dominant contribution to the visible sector
width \cite{Rosa:2022tch},
\begin{equation}
    \Gamma
    \simeq
    \Gamma_{\phi\rightarrow HH}.
\end{equation}

If the inflaton is also sufficiently heavy compared with the dark
Higgs,
\begin{equation}
    m_{\phi}>2M_{H'},
\end{equation}
the same 2-body channel opens in the dark sector. Its width is
\begin{equation}
    \Gamma'_{\phi\rightarrow H'H'}
    =
    \frac{g'^{\,2}v_{\phi}^{2}}
         {32\pi m_{\phi}}
    \left(
        1-\frac{4M_{H'}^{2}}{m_{\phi}^{2}}
    \right)^{1/2}.
    \label{eq:2-body-width-dark}
\end{equation}
The resulting reheating asymmetry is therefore
\begin{equation}
    \epsilon_{2}
    \equiv
    \frac{\Gamma'_{\phi\rightarrow H'H'}}
         {\Gamma_{\phi\rightarrow HH}}
    =
    \left(
        \frac{g'}{g}
    \right)^{2}
    \frac{
        \left(
            1-4M_{H'}^{2}/m_{\phi}^{2}
        \right)^{1/2}
    }{
        \left(
            1-4M_H^{2}/m_{\phi}^{2}
        \right)^{1/2}
    }.
    \label{eq:epsilon-2-body-general}
\end{equation}
Since $M_H\ll m_{\phi}$, this becomes
\begin{equation}
    \epsilon_{2}
    \simeq
    \left(
        \frac{g'}{g}
    \right)^{2}
    \left(
        1-\frac{4M_{H'}^{2}}{m_{\phi}^{2}}
    \right)^{1/2}.
    \label{eq:epsilon-2-body}
\end{equation}
Away from the dark Higgs threshold, the two sectors are therefore
reheated with comparable efficiencies if $g'\sim g$. A sizeable
asymmetry in this regime requires the dark decay to occur close to the
kinematic threshold.

\subsection{3-body decay}
\label{sec:threebodyr}

We next consider the intermediate mass hierarchy
\begin{equation}
    M'_H < m_\phi < 2M'_H ,
\end{equation}
in which the single-resonant 3-body topology can become
kinematically accessible. Provided that
$m_\phi>M'_H+2m'_f$, one dark Higgs boson can be produced on-shell,
while the second remains virtual and subsequently decays into a dark
fermion-antifermion pair,
\begin{equation}
    \phi(p)
    \longrightarrow H'(k)+H'^{*}(q)
    \longrightarrow H'(k)+f'(p_1)+\bar f'(p_2).
    \label{eq:threebody_process}
\end{equation}
Defining
\begin{equation}
    q=p_1+p_2,
    \qquad
    s\equiv q^2 ,
\end{equation}
the kinematically allowed range for the invariant mass of the fermion pair is
\begin{equation}
    4m_{f'}^2
    \leq s
    \leq
    \left(m_\phi-M_{H'}\right)^2 .
    \label{eq:threebody_slimits}
\end{equation}

We assume that the dark fermion acquires its mass from the dark Higgs
symmetry-breaking scale $v_{H'}$. Motivated by
$v_{H'}\sim M_{H'}$, for the numerical analysis we impose the
benchmark relation
\begin{equation}
    m_{f'} = y'M_{H'} .
    \label{eq:threebody_mf}
\end{equation}
Here, convention- and mixing-dependent factors are absorbed into the
effective coupling $y'$.

It is therefore important to retain the final-state fermion mass when
varying $y'$. Introducing
\begin{equation}
    z\equiv \frac{m_\phi}{M_{H'}},
\end{equation}
the condition $m_\phi \geq M_{H'}+2m_{f'}$ implies
\begin{equation}
    0<y'<\frac{z-1}{2},
    \qquad 1<z<2 .
    \label{eq:threebody_ybound}
\end{equation}

The 3-body phase-space measure factorises as
\begin{equation}
    d\Phi_3(p;k,p_1,p_2)
    =
    \frac{ds}{2\pi}\,
    d\Phi_2(p;k,q)\,
    d\Phi_2(q;p_1,p_2).
    \label{eq:threebody_phase_space}
\end{equation}
The corresponding matrix element contains one virtual dark Higgs propagator.
Close to the upper boundary of the 3-body regime,
$m_\phi\rightarrow 2M_{H'}^{-}$, the virtual state approaches its mass shell.
We therefore retain its finite decay width,
\begin{equation}
    \frac{1}{(s-M_{H'}^2)^2}
    \longrightarrow
    \frac{1}
    {(s-M_{H'}^2)^2+M_{H'}^2\Gamma_{H'}^2}.
    \label{eq:threebody_BW}
\end{equation}
After summing over the final-state fermion spins, the differential decay
width becomes
\begin{equation}
\frac{d\Gamma'_3}{ds}
=
\frac{g'^2 y'^2 v_\phi^2}
     {128\pi^3 m_\phi^3}
\,
\frac{
s\,\lambda^{1/2}(m_\phi^2,M_{H'}^2,s)
}{
(s-M_{H'}^2)^2+M_{H'}^2\Gamma_{H'}^2
}
\left(
1-\frac{4m_{f'}^2}{s}
\right)^{3/2},
\label{eq:threebody_dGamma}
\end{equation}
where $\lambda$ denotes the K\"all\'en function.

It is convenient to introduce
\begin{equation}
    r\equiv\frac{M_{H'}}{m_\phi}=\frac{1}{z},
    \qquad
    x\equiv\frac{s}{m_\phi^2},
    \qquad
    \gamma\equiv\frac{\Gamma_{H'}}{M_{H'}} .
\end{equation}
The total width can then be written as
\begin{equation}
    \Gamma'_3
    =
    \frac{g'^2 y'^2 v_\phi^2}
         {128\pi^3 m_\phi}
    F_3(r,\gamma,y'),
    \label{eq:threebody_Gamma}
\end{equation}
where
\begin{equation}
\begin{split}
F_3(r,\gamma,y')
={}&
\int_{4y'^2r^2}^{(1-r)^2} dx\,
\frac{
x\,\lambda^{1/2}(1,r^2,x)
}{
(x-r^2)^2+r^4\gamma^2
}
\times
\left(
1-\frac{4y'^2r^2}{x}
\right)^{3/2}.
\end{split}
\label{eq:threebody_F3}
\end{equation}

For a single fermion species, assuming that $H'\to f'\bar f'$ dominates
the total width of the intermediate state, we use
\begin{equation}
    \Gamma_{H'}
    \simeq
    \Gamma_{H'\to f'\bar f'}
    =
    \frac{y'^2 M_{H'}}{8\pi}
    \left(1-4y'^2\right)^{3/2}.
    \label{eq:dark Higgs-width}
\end{equation}
and hence
\begin{equation}
    \gamma(y')
    =
    \frac{y'^2}{8\pi}
    \left(1-4y'^2\right)^{3/2}.
    \label{eq:threebody_gamma}
\end{equation}
Throughout this analysis, we consider a single colour-singlet Dirac
fermion species, corresponding to $N_c=1$, and use the on-shell value
of $\Gamma_{H'}$ as a constant in the Breit-Wigner propagator. This
prescription regulates the resonant enhancement as $z\to2^-$. Allowing
$\Gamma_{H'}$ to depend on the virtuality could affect the resulting
line shape and hence the precise numerical bounds.
In the immediate vicinity of the pair-production threshold, however,
the separation between the single-resonant 3-body and on-shell
2-body descriptions is no longer uniform.
In Sec.~\ref{sec:neff}, we therefore use an approximate finite-width treatment of the four-fermion final state to interpolate across $z=2$.

The reheating asymmetry in the 3-body regime is
\begin{equation}
    \epsilon_3
    \equiv
    \frac{\Gamma'_3}{\Gamma_{\phi\rightarrow HH}} .
\end{equation}
For $M_H\ll m_\phi$, this yields
\begin{equation}
    \epsilon_3(z,y')
    \simeq
    \frac{1}{4\pi^2}
    \left(\frac{g'}{g}\right)^2
    y'^2
    F_3\left(
    \frac{1}{z},\gamma(y'),y'
    \right).
    \label{eq:threebody_epsilon}
\end{equation}

To illustrate the largest reheating asymmetry attainable for each mass ratio,
we maximise over the kinematically allowed Yukawa coupling,
\begin{equation}
\epsilon_3^{\rm max}(z)
=
\frac{1}{4\pi^2}
\left(\frac{g'}{g}\right)^2
\max_{0<y'<(z-1)/2}
\left[
y'^2
F_3\left(
\frac{1}{z},\gamma(y'),y'
\right)
\right].
\label{eq:threebody_envelope}
\end{equation}
Notice that the kinematic upper limit
$y'=(z-1)/2$ does not maximise the physical decay rate, since it corresponds
to the endpoint $m_\phi=M_{H'}+2m_{f'}$, where the available phase space
vanishes. The maximum therefore occurs at an intermediate value of $y'$.

The 3-body rate is strongly suppressed near $z=1$, where the available
phase space closes, and grows rapidly as the 2-body threshold is
approached from below. In this sense, the 3-body channel occupies the
intermediate region between the strongly suppressed 4-body process and
the fully on-shell 2-body decay. The corresponding implications for
$\Delta N_{\rm eff}$ are discussed in Sec.~\ref{sec:neff}.

\subsection{4-body decay}
\label{subsec:4-body}

Finally, we consider the regime
\begin{equation}
    m_\phi < M_{H'},
    \label{eq:fourbody_hierarchy}
\end{equation}
in which neither of the dark Higgs bosons can be produced on shell.
The leading inflaton decay into the dark sector then proceeds through
two virtual dark Higgs bosons,
\begin{equation}
    \phi(p)
    \longrightarrow
    H'^{*}(q_1)+H'^{*}(q_2)
    \longrightarrow
    f'(p_1)+\bar f'(p_2)+f'(p_3)+\bar f'(p_4).
    \label{eq:fourbody_process}
\end{equation}
We define
\begin{equation}
    q_1=p_1+p_2,
    \qquad
    q_2=p_3+p_4,
\end{equation}
and introduce the invariant masses
\begin{equation}
    s_1\equiv q_1^2,
    \qquad
    s_2\equiv q_2^2.
\end{equation}
Thus, $\sqrt{s_1}$ and $\sqrt{s_2}$ correspond to the invariant
masses of the two fermion-antifermion pairs. Since both dark Higgs
bosons are virtual, $s_1$ and $s_2$ vary independently over the
kinematically allowed 4-body phase space.

As in the 3-body regime, we retain the dark fermion mass,
\begin{equation}
    m_{f'} = y'M_{H'}.
    \label{eq:mf_fourbody}
\end{equation}
For a single dark fermion species, the complete four-fermion amplitude
also contains the exchange contribution obtained by interchanging the
two identical antifermions, together with the corresponding interference
term. In the following, as in the factorised treatment adopted in the
original calculation, we retain one fermion-antifermion pairing and
include the identical-particle statistical factor separately. The
resulting expression should therefore be understood as the factorised
4-body contribution rather than the complete identical-particle
amplitude.
For the factorised contribution considered here, the spin-summed
squared matrix element is
\begin{equation}
    \sum_{\rm spins} |\mathcal M_4|^2
    =
    4g'^2y'^4v_\phi^2
    \frac{
    (s_1-4m_{f'}^2)(s_2-4m_{f'}^2)
    }{
    (s_1-M_{H'}^2)^2
    (s_2-M_{H'}^2)^2
    }.
    \label{eq:M4_squared}
\end{equation}
Following the identical-particle convention adopted in the original
4-body calculation, we include the statistical factor
\begin{equation}
    S=\frac{1}{2!2!}=\frac14
\end{equation}
for the two identical fermions and two identical antifermions.

The 4-body phase-space measure can be factorised as
\begin{equation}
    d\Phi_4
    =
    \frac{ds_1}{2\pi}
    \frac{ds_2}{2\pi}
    d\Phi_2(p;q_1,q_2)
    d\Phi_2(q_1;p_1,p_2)
    d\Phi_2(q_2;p_3,p_4).
\end{equation}
After integration over the angular variables,
\begin{equation}
    d\Phi_4
    =
    \frac{ds_1\,ds_2}{2048\pi^5}
    \frac{\lambda^{1/2}(m_\phi^2,s_1,s_2)}{m_\phi^2}
    \prod_{i=1}^{2}
    \left(
        1-\frac{4m_{f'}^2}{s_i}
    \right)^{1/2}.
    \label{eq:fourbody_phase_space}
\end{equation}
Combining the matrix element and phase space gives
\begin{align}
    \Gamma'_4
    =
    \frac{g'^2y'^4v_\phi^2}
         {4096\pi^5m_\phi^3}
    \int_{\mathcal R} ds_1\,ds_2\,
    \frac{
    s_1s_2\,
    \lambda^{1/2}(m_\phi^2,s_1,s_2)
    }{
    (s_1-M_{H'}^2)^2
    (s_2-M_{H'}^2)^2
    }
    \left(
        1-\frac{4m_{f'}^2}{s_1}
    \right)^{3/2}
    \left(
        1-\frac{4m_{f'}^2}{s_2}
    \right)^{3/2}.
    \label{eq:Gamma4_massive}
\end{align}
The integration region is
\begin{equation}
    4m_{f'}^2
    \leq s_1
    \leq
    (m_\phi-2m_{f'})^2,
\end{equation}
and, for fixed $s_1$,
\begin{equation}
    4m_{f'}^2
    \leq s_2
    \leq
    (m_\phi-\sqrt{s_1})^2.
\end{equation}
A detailed derivation of Eq.~\eqref{eq:Gamma4_massive}, including the
factorisation of the 4-body phase space and the treatment of the
massive final-state fermions, is given in
Appendix~\ref{app:4-body}.

It is useful to introduce
\begin{equation}
    z\equiv\frac{m_\phi}{M_{H'}},
    \qquad
    \mu\equiv\frac{m_{f'}}{m_\phi}
    =\frac{y'}{z},
    \qquad
    x_i\equiv\frac{s_i}{m_\phi^2}.
    \label{eq:fourbody_dimensionless}
\end{equation}
Kinematic accessibility of the four-fermion final state requires
\begin{equation}
    m_\phi>4m_{f'},
\end{equation}
and therefore
\begin{equation}
    0<\mu<\frac14,
    \qquad
    0<y'<\frac{z}{4}.
    \label{eq:fourbody_kinematic_bound}
\end{equation}

Although the 4-body expression above is introduced for $z<1$, the
four-fermion final state also remains kinematically allowed for $z>1$.
In particular, when
\begin{equation}
    y' > \frac{z-1}{2},
\end{equation}
there is insufficient phase space to produce even one dark Higgs on
shell together with the remaining fermion pair. Both dark Higgs
propagators are then necessarily off-shell, and the same factorised
4-body expression can be used in this non-resonant region. We will
make use of this continuation when discussing the transition across
$z=1$ in Sec.~\ref{sec:neff}.

In terms of the dimensionless variables, the decay width becomes
\begin{equation}
    \Gamma'_4
    =
    \frac{g'^2y'^4v_\phi^2}
         {4096\pi^5m_\phi}
    z^8J_4(z,\mu),
    \label{eq:Gamma4_J4}
\end{equation}
where
\begin{align}
    J_4(z,\mu)
    =
    \int_{4\mu^2}^{(1-2\mu)^2}\!\!\!\!dx_1
    \int_{4\mu^2}^{(1-\sqrt{x_1})^2}\!\!\!\!dx_2
    \,
    \frac{
    x_1x_2\lambda^{1/2}(1,x_1,x_2)
    }{
    (1-z^2x_1)^2(1-z^2x_2)^2
    }
    \left(
        1-\frac{4\mu^2}{x_1}
    \right)^{3/2}
    \left(
        1-\frac{4\mu^2}{x_2}
    \right)^{3/2}.
    \label{eq:J4}
\end{align}

Using the visible-sector 2-body width,
\begin{equation}
    \Gamma_{\phi\rightarrow HH}
    \simeq
    \frac{g^2v_\phi^2}{32\pi m_\phi},
\end{equation}
the reheating asymmetry in the 4-body regime is
\begin{equation}
    \epsilon_4(z,\mu)
    =
    \frac{1}{128\pi^4}
    \left(\frac{g'}{g}\right)^2
    z^{12}\mu^4J_4(z,\mu).
    \label{eq:epsilon4_full}
\end{equation}

To determine the largest reheating asymmetry attainable for each
mass ratio, we maximise over the kinematically allowed fermion mass,
or equivalently over the dark Yukawa coupling,
\begin{equation}
    \epsilon_4^{\rm max}(z)
    =
    \frac{1}{128\pi^4}
    \left(\frac{g'}{g}\right)^2
    z^{12}
    \max_{0<\mu<1/4}
    \left[
        \mu^4J_4(z,\mu)
    \right].
    \label{eq:epsilon4_max}
\end{equation}
For $z<1$, the maximisation is performed over the full kinematically
allowed interval $0<\mu<1/4$. For the double-off-shell continuation
into the intermediate region $1<z<2$, however, one must additionally
require that the single-resonant configuration be kinematically
inaccessible. Using $\mu=y'/z$, this corresponds to
\begin{equation}
    \frac{z-1}{2z}<\mu<\frac14 .
\end{equation}
We use this restricted interval when evaluating the double-off-shell
contribution above $z=1$.

Notice that the kinematic endpoint $\mu=1/4$, corresponding to
$y'=z/4$, does not maximise the decay rate. At this point
$m_\phi=4m_{f'}$, so that the 4-body phase space closes and
$\Gamma'_4$ vanishes. The physical maximum therefore occurs at an
intermediate value of the Yukawa coupling.

As a consistency check, in the strongly hierarchical and massless
limits, $z\ll1$ and $\mu\rightarrow0$,
\begin{equation}
    J_4(z,\mu)\longrightarrow\frac{1}{720},
\end{equation}
and hence
\begin{equation}
    \Gamma'_4
    \longrightarrow
    \frac{g'^2y'^4v_\phi^2}
         {2949120\,\pi^5m_\phi}
    \left(\frac{m_\phi}{M_{H'}}\right)^8.
    \label{eq:Gamma4_deep_offshell}
\end{equation}
The characteristic strong off-shell suppression of the 4-body
channel is therefore recovered in the hierarchical limit. The
cosmological implications of this suppression are discussed in
Secs.~\ref{sec:cosmology} and \ref{sec:neff}.

\section{Cosmological implications of the 4-body regime}
\label{sec:cosmology}

We now examine the thermal history of the dark sector after reheating, focusing on
the strongly hierarchical 4-body decay regime,
\begin{equation}
    M_{\rm SUSY} \ll m_\phi < M_{H'} < M'_{\rm SUSY},
    \label{eq:hierarchical-regime}
\end{equation}
which naturally arises in the two-sector framework of
Ref.~\cite{Rosa:2022tch}. In this regime, the inflaton can decay into
visible Higgs bosons, whereas both dark Higgs bosons are necessarily
off-shell. The leading dark sector decay is therefore the 4-body
process discussed in Sec.~\ref{subsec:4-body}.

The corresponding energy-density and temperature evolution in the
3- and 2-body regimes is obtained from the general solutions of
Sec.~\ref{sec:reheating} by using the appropriate dark sector partial width
$\Gamma'_i$ and the corresponding total inflaton width
$\Gamma_\phi=\Gamma+\Gamma'_i$, with $i=3,2$, respectively.
We do not repeat these channel-independent solutions
here; the three decay regimes are compared through
$\Delta N_{\rm eff}$ in Sec.~\ref{sec:neff}. Unless otherwise stated, the numerical
results presented in this section consider the reference values $T_{\rm RH}=100~{\rm GeV}$,
$V(\phi_e)^{1/4}=10^{16}~{\rm GeV}$, $g_*=g_*'=100$, $g'=g$, and
$M_{H'}/M'_{\rm SUSY}=0.1$, the latter corresponding to the SUSY scale setting the electroweak scale up to loop factors.

\subsection{Thermal evolution of the dark sector}
\label{subsec:dark-thermal-evolution}
For the 4-body channel, the reheating asymmetry obtained in
Sec.~\ref{subsec:4-body} is
\begin{equation}
    \epsilon_4(z,\mu)
    =
    \frac{1}{128\pi^4}
    \left(\frac{g'}{g}\right)^2
    z^{12}\mu^4 J_4(z,\mu),
    \qquad
    z\equiv\frac{m_\phi}{M_{H'}},
    \qquad
    \mu\equiv\frac{m_{f'}}{m_\phi}.
    \label{eq:epsilon4_cosmo}
\end{equation}
For each value of $z$, the largest energy transfer to the dark sector
is obtained by maximising over the kinematically allowed interval
$0<\mu<1/4$,
\begin{equation}
    \epsilon_4^{\rm max}(z)
    =
    \frac{1}{128\pi^4}
    \left(\frac{g'}{g}\right)^2
    z^{12}
    \max_{0<\mu<1/4}
    \left[
        \mu^4 J_4(z,\mu)
    \right].
    \label{eq:epsilon4max_cosmo}
\end{equation}
The reheating process is therefore naturally highly asymmetric in the
4-body regime, with the energy transferred to the dark sector
becoming increasingly suppressed as the hierarchy between the inflaton
and dark Higgs masses increases.

Using the temperature relation derived in Sec.~\ref{subsec:temperature-asymmetry}, the corresponding
maximal dark-to-visible temperature ratio is
\begin{equation}
    \left(\frac{T'}{T}\right)_{\rm max}
    =
    \left[\epsilon_4^{\rm max}(z)\right]^{1/4}
    \left[
        \frac{g_\ast(T)}
             {g'_\ast(T')}
    \right]^{1/4}.
    \label{eq:Tratio_fourbody}
\end{equation}
In the strongly hierarchical regime, $z\ll1$, the full result
approaches the deep off-shell limit and the characteristic scaling
$\epsilon_4^{\rm max}\propto z^{12}$ is recovered. Consequently,
\begin{equation}
    \frac{T'}{T}\propto \left({m_\phi\over m_{H'}}\right)^3,
    \qquad m_\phi\ll M_{H'},
    \label{eq:Tratio_scaling}
\end{equation}
up to the dependence on the number of relativistic degrees of freedom and on
the coupling ratio $g'/g$. Thus, the strong suppression of the
dark sector temperature found in the hierarchical regime is preserved
when the finite fermion mass and the full virtual-scalar propagators
are included.

\begin{figure}[H]
    \centering
    \includegraphics[width=0.75\textwidth]
    {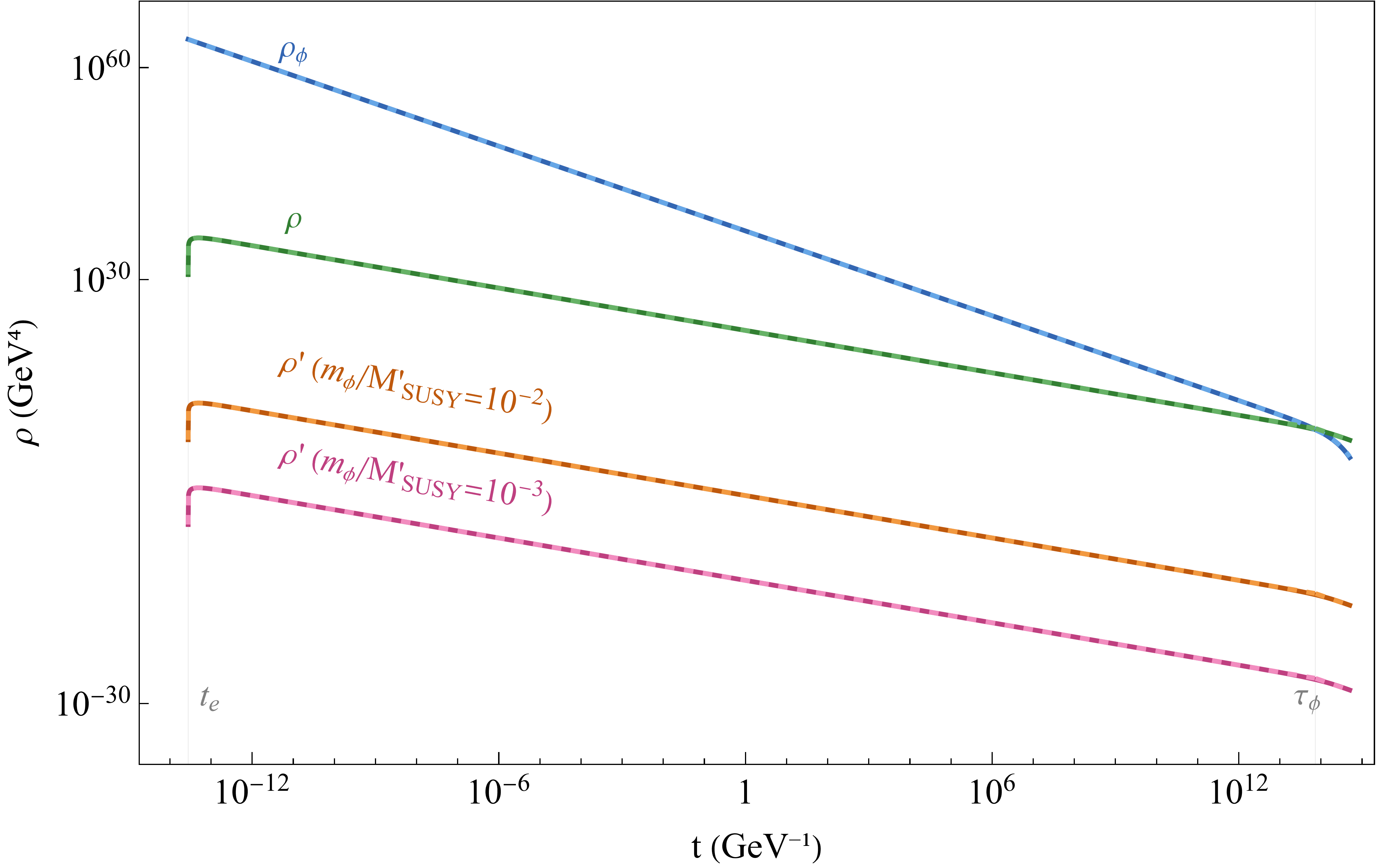}
    \caption{
   Evolution of the inflaton energy density, $\rho_\phi$, the
visible-sector radiation density, $\rho$, and the dark sector radiation
density, $\rho'$, for $m_\phi/M'_{\rm SUSY}=10^{-2}$ and
$m_\phi/M'_{\rm SUSY}=10^{-3}$. We take $g'=g$ and
$M'_H/M'_{\rm SUSY}=0.1$. The upper and lower dark sector curves
correspond to $m_\phi/M'_{\rm SUSY}=10^{-2}$ and $10^{-3}$,
respectively. Solid curves represent the numerical solutions, while
dashed curves correspond to the analytical approximations.
    }
    \label{fig:energy-densities-hierarchies}
\end{figure}

The corresponding evolution of the energy densities is shown in
Figure~\ref{fig:energy-densities-hierarchies}.
The inflaton and the visible sector densities are essentially unaffected by the choice of
dark sector mass hierarchy, while the dark radiation density decreases
rapidly as $m_\phi/M'_{\rm SUSY}$ becomes smaller.

The dark sector reheating and maximum temperatures inherit the same
strong dependence on the mass hierarchy. The dark sector reheating
temperature can be written as
\begin{equation}
    T'_{\rm RH}
    \simeq
    T_{\rm RH}
    \left[
        \epsilon_4^{\rm max}(z)
        \frac{g_\ast(T_{\rm RH})}
             {g'_\ast(T'_{\rm RH})}
    \right]^{1/4}.
    \label{eq:TRHprime}
\end{equation}
The maximum temperature reached by the dark radiation bath follows
from Eq.~(\ref{eq:Tmaxdark}). Using
$\Gamma'_4=\epsilon_4^{\rm max}(z)\Gamma$, one obtains
\begin{equation}
    T'_{\rm max}
    \simeq
    \left[
        \frac{30\sqrt{3}}
        {4\pi^2 g'_*(T'_{\rm max})}
        \left(\frac{3}{8}\right)^{3/5}
    \right]^{1/4}
    \left[
        M_{\rm Pl}\Gamma\,
        \epsilon_4^{\rm max}(z)
    \right]^{1/4}
    V(\phi_e)^{1/8}.
\end{equation}
In the deep off-shell regime, both temperatures therefore retain the
characteristic scaling
\begin{equation}
    T'_{\rm RH},\,T'_{\rm max}\propto z^3.
\end{equation}
%

\begin{figure}[H]
    \centering
    \includegraphics[width=0.75\textwidth]
    {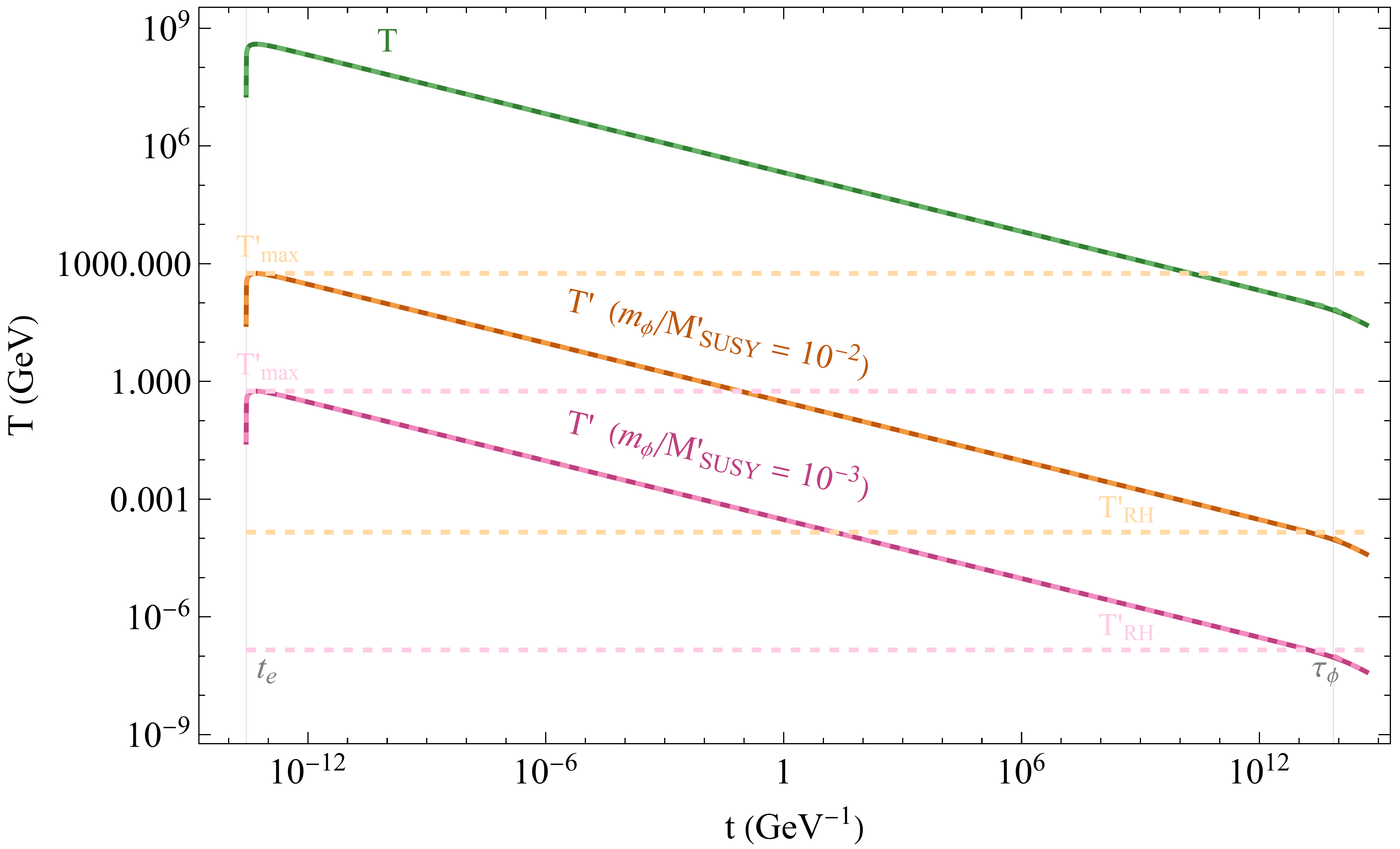}
    \caption{
   Evolution of the visible-sector temperature, $T$, and of the
dark sector temperature, $T'$, for $m_\phi/M'_{\rm SUSY}=10^{-2}$ and
$m_\phi/M'_{\rm SUSY}=10^{-3}$. We take $g'=g$,
$M'_H/M'_{\rm SUSY}=0.1$, and $g_*=g_*'=100$. The upper and lower
dark sector curves correspond to $m_\phi/M'_{\rm SUSY}=10^{-2}$ and
$10^{-3}$, respectively. Solid curves represent the numerical solutions,
while dashed curves correspond to the analytical approximations.
    }
    \label{fig:temperatures-hierarchies}
\end{figure}

The resulting temperature evolution is displayed in
Figure~\ref{fig:temperatures-hierarchies}. Increasing the hierarchy
between the inflaton and dark supersymmetry-breaking scales suppresses
both the maximum temperature and the subsequent temperature of the
dark radiation bath.

These results show that the 4-body decay regime leads naturally to
a dark sector that is much colder than the visible one. The maximum temperature reached during reheating is particularly important, since it determines which dark sector processes and phase transitions could
have taken place.

\subsection{Dark QCD crossover}
\label{subsec:dark-qcd}

We next investigate whether the dark sector ever reached temperatures
above its QCD confinement scale. This discussion assumes that the strongly interacting dark degrees of
freedom establish approximate thermal equilibrium sufficiently rapidly
for $T'$ to be interpreted as a physical temperature of the dark QCD
plasma. As discussed in
Section~\ref{subsec:dark-abundance}, the dark QCD scale,
$\Lambda'_{\rm QCD}$, depends on the dark supersymmetry-breaking scale
and on the corresponding quark-mass thresholds.

Since the dark sector temperature decreases after reaching its maximum, dark QCD deconfinement can only occur if
\begin{equation}
    T'_{\rm max}
    >
    \Lambda'_{\rm QCD}.
    \label{eq:dark-qcd-transition-condition}
\end{equation}
If instead
\begin{equation}
    T'_{\rm max}
    <
    \Lambda'_{\rm QCD},
    \label{eq:no-dark-qcd-transition}
\end{equation}
the dark sector never enters a deconfined quark-gluon phase. In that case, if coloured dark sector degrees of freedom are populated,
they must hadronise rapidly into dark bound states rather than forming
a deconfined thermal bath of dark quarks and gluons.

\begin{figure}[H]
    \centering
    \includegraphics[width=0.55\textwidth]
    {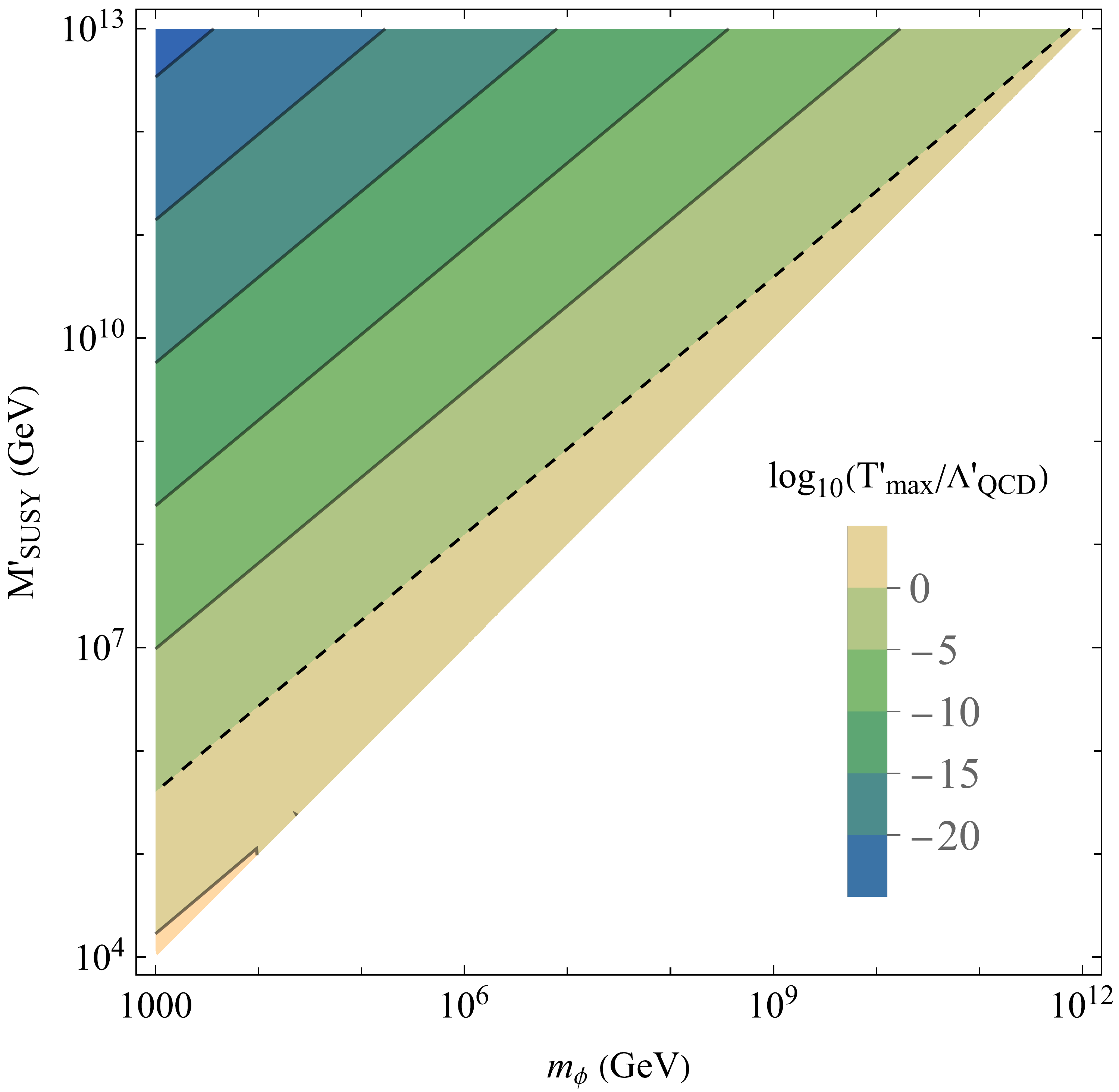}
    \caption{
    Ratio $T'_{\max}/\Lambda'_{\rm QCD}$ as a function of the
inflaton mass, $m_\phi$, and the dark supersymmetry-breaking scale,
$M'_{\rm SUSY}$. The dashed line corresponds to
$T'_{\max}=\Lambda'_{\rm QCD}$ and separates the regions with and
without a dark QCD deconfinement transition. We take $m'_t/M'_{\rm SUSY}\simeq M'_H/M'_{\rm SUSY}=0.1$. Only the region $m_\phi<M_{H'}=0.1M'_{\rm SUSY}$, in which the
double-off-shell 4-body description applies, is shown.
    }
    \label{fig:dark-qcd}
\end{figure}

The ratio between the maximum dark sector temperature and the dark QCD
scale is shown in Figure~\ref{fig:dark-qcd}. The dashed contour
corresponds to
\begin{equation}
    T'_{\rm max}
    =
    \Lambda'_{\rm QCD},
\end{equation}
and separates the regions with and without a dark QCD crossover.

For most of the parameter region corresponding to
\begin{equation}
    M'_{\rm SUSY}\gg m_\phi,
\end{equation}
the maximum dark sector temperature remains below the dark confinement
scale. The dark sector, therefore, does not undergo a QCD phase
transition in the strongly hierarchical regime considered here.

A thermal evolution more similar to that of the visible sector remains
possible when $M'_{\rm SUSY}$ lies much closer to the inflaton mass. In this region, the suppression of the inflaton decay width is weaker and the dark sector may reach temperatures above
$\Lambda'_{\rm QCD}$.

\subsection{Dark weak interactions}
\label{subsec:dark-weak-interactions}

We finally consider whether weak interactions in the dark sector could
have reached thermal equilibrium. In the visible sector, neutrinos are
kept in equilibrium by processes such as
\begin{equation}
    \nu_e+e^-
    \longleftrightarrow
    \nu_e+e^-,
    \qquad
    e^-+e^+
    \longleftrightarrow
    \nu_e+\bar{\nu}_e.
    \label{eq:visible-weak-processes}
\end{equation}
At temperatures well below the weak gauge-boson masses, these processes
are described by the effective Fermi theory. As the temperature decreases,
their interaction rates rapidly become inefficient compared with the Hubble
expansion rate, while electron-positron annihilation is additionally suppressed
once the temperature drops below the electron mass.

Since the two sectors are assumed to share the same gauge structure,
analogous processes are expected to occur in the dark sector. Their cross-section may be parametrised as
\begin{equation}
    \sigma'
    =
    {G'_F}^{\,2}T'^{\,2},
    \label{eq:dark-weak-cross-section}
\end{equation}
where the dark Fermi constant is
\begin{equation}
    G'_F
    =
    \frac{\sqrt{2}}{8}
    \left(
        \frac{g'_W}{m'_W}
    \right)^2
    \simeq
    \frac{\sqrt{2}}{8}
    \left(
        \frac{g'_W}{M'_{H}}
    \right)^2.
    \label{eq:dark-fermi-constant}
\end{equation}
In the last estimate, we take the dark weak gauge boson mass to be of
the same order as the dark Higgs mass, $m'_W\sim M_{H'}$. For the
benchmark values considered below, the center-of-mass energy of the
primary particles produced in inflaton decays is bounded by
$\sqrt{s_{\rm cm}}\lesssim m_\phi$. Moreover,
$m_\phi/m'_W\sim 10^{-1}$ and $10^{-2}$ for the two representative mass
ratios considered below. The effective Fermi description is therefore
self-consistent both for the thermal estimate and for the initially
non-thermal decay products.
Assuming $g'_W/g_W\sim\mathcal{O}(1)$ and taking
$M'_H/M'_{\rm SUSY}=0.1$, as above, this gives approximately 
\begin{equation}
    G'_F
    \sim
    10^{-17}\ {\rm GeV}^{-2}
    \left(
        \frac{10^{8}\ {\rm GeV}}
             {M'_{\rm SUSY}}
    \right)^2.
    \label{eq:dark-fermi-constant-intermediate-scale}
\end{equation}
where we considered the reference value
$M'_{\rm SUSY}=10^8\,\mathrm{GeV}$ yielding dark protons/neutrons with $m_{p'}\simeq m_{n'}\simeq 1$ TeV, which may potentially lead to the formation of planetary-mass dark neutron stars as discussed in \cite{Litterer:2025DNS}.

For relativistic particles, the visible sector weak interaction rate is
\begin{equation}
\Gamma_\nu \simeq n\sigma =
\left(g_{e^-}+g_{\nu_e}\right)
\frac{3\zeta(3)}{4\pi^2}
G_F^2 T^5 .
 \label{eq:visible-neutrino-rate}
\end{equation}

\begin{figure}[H]
    \centering
    \includegraphics[width=0.75\textwidth]
    {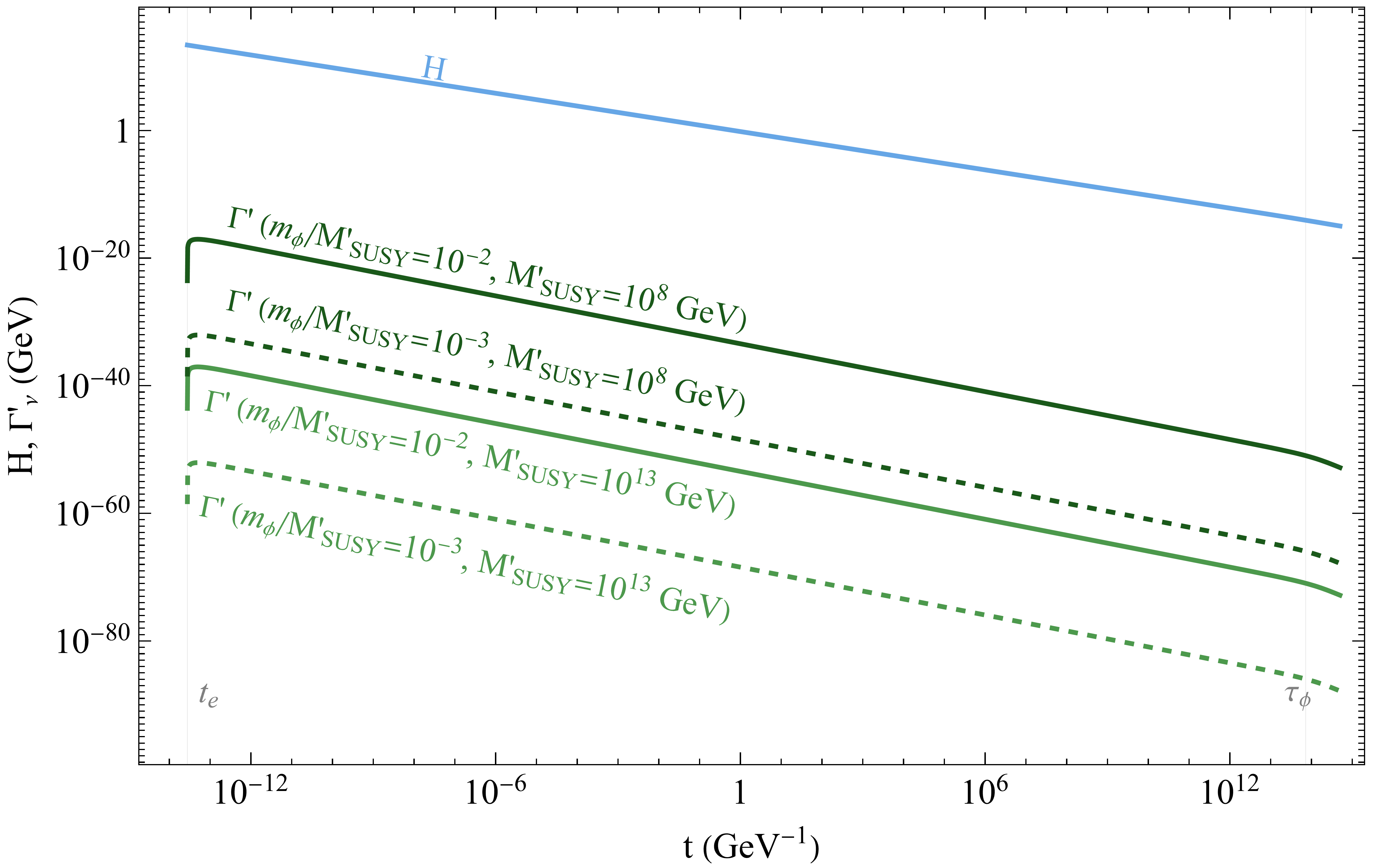}
    \caption{
   Evolution of the Hubble parameter and of the thermal estimate of the
dark weak-interaction rate.
Solid and dashed curves correspond to the two inflaton-to-SUSY mass ratios,
while darker and lighter curves correspond to the two values of
$M'_{\rm SUSY}$}
    \label{fig:dark-weak-interactions}
\end{figure}

Dark sector particles, in particular neutrinos, may be substantially heavier, in which case the
relativistic approximation would not be valid, and dark neutrino decoupling would occur even earlier. Since the dark sector distribution need not be thermal, a complete thermalization analysis would require following the non-thermal phase-space distributions of the particles produced in inflaton decays.
As a benchmark, we assume relativistic dark neutrinos and evaluate the
interaction rate using the equilibrium number density associated with
the dark sector temperature $T'$.
\begin{equation}
\Gamma'_\nu \simeq n'\sigma' =
\left(g_{e'^-}+g_{\nu'_e}\right)
\frac{3\zeta(3)}{4\pi^2}
G_F'^2 T'^5 .
\label{eq:dark-neutrino-rate}
\end{equation}
Thermal equilibrium requires the interaction rate to exceed the Hubble
expansion rate,
\begin{equation}
    \Gamma'_\nu
    \gtrsim
    H.
    \label{eq:dark-weak-equilibrium-condition}
\end{equation}
Figure~\ref{fig:dark-weak-interactions} compares these quantities for the representative mass ratios and two different values of the dark
SUSY breaking scale.

For all the benchmark values shown, the thermal-rate estimate gives $\Gamma'_\nu \ll H$, so that even within the optimistic relativistic thermal approximation, dark weak interactions are therefore unable to maintain thermal equilibrium. We interpret
Fig.~\ref{fig:dark-weak-interactions} as indicating that dark weak thermalization is disfavoured for the representative benchmarks considered here, in line with the estimates in \cite{Litterer:2025DNS}. This suggests a
cosmological history markedly different from that of the visible
sector, where weak interactions are in equilibrium at high temperatures
and neutrinos decouple only around $T\sim 1~\mathrm{MeV}$.



\section{Dark radiation constraints}
\label{sec:neff}

We now discuss the contribution of the dark radiation bath to the effective number of relativistic neutrino species, $\Delta N_{\rm eff}$. Additional relativistic degrees of freedom are constrained by
cosmological observations, including BBN, CMB, and BAO measurements,
and therefore provide a possible probe of asymmetric reheating
\cite{ParticleDataGroup:2024cfk,Planck:2018vyg, Abazajian:2022CMB-S4,Goldstein:2026iuu}.

The resulting contribution depends both on the temperature hierarchy between the two
sectors and on the number of dark degrees of freedom that remain
relativistic at the relevant cosmological epochs.

\subsection{Dark neutrinos and relativistic degrees of freedom}
\label{subsec:dark-neutrinos}

The relativistic particle content of the dark sector is uncertain,
since the masses of its particles may be substantially larger than
their visible-sector counterparts. In particular, dark neutrinos may
already be non-relativistic at reheating.

Assuming the existence of right-handed neutrinos and a Majorana mass
scale common to both sectors, the seesaw mechanism gives \cite{Minkowski:1977sc, Mohapatra:1979ia}:
\begin{equation}
    m_{\nu}
    =
    \frac{y^{2}v_{H}^{2}}{M_{\nu}},
    \qquad
    m'_{\nu}
    =
    \frac{y'^{\,2}v_{H}'^{\,2}}{M_{\nu}},
    \label{eq:visible-dark-seesaw}
\end{equation}
where $y$ and $y'$ are the visible- and dark sector Yukawa couplings,
$v_H$ and $v'_H$ are the corresponding Higgs vacuum expectation
values, and $M_\nu$ is the Majorana mass.

Taking
\begin{equation}
    v'_H\sim M'_H,
    \qquad
    \frac{y'}{y}\sim\mathcal{O}(1),
\end{equation}
the dark neutrino mass can be related to the visible one as
\begin{equation}
    m'_{\nu}
    =
    \left(
        \frac{y'}{y}
    \right)^{2}
    \left(
        \frac{M'_{H}}{v_{H}}
    \right)^{2}
    m_{\nu}.
    \label{eq:dark-neutrino-mass}
\end{equation}
Taking $m_\nu=10^{-2}\,\mathrm{eV}$, $v_H=246\,\mathrm{GeV}$, and
$M'_H=0.1\,M'_{\rm SUSY}$, for the two representative dark
supersymmetry-breaking scales considered in Section~\ref{subsec:dark-weak-interactions}, this gives
\begin{align}
    m'_{\nu}
    &\simeq
    20\ {\rm MeV}
    \left(
        \frac{y'}{y}
    \right)^2
    \left(
        \frac{M'_{\rm SUSY}}
             {10^{8}\ {\rm GeV}}
    \right)^2.
    \label{eq:dark-neutrino-mass-intermediate}
\end{align}

Whether dark neutrinos contribute to the relativistic dark sector
degrees of freedom depends on both the maximum temperature reached by
the dark sector and on their production history. For the benchmark
considered below, we assume that dark neutrinos do not provide an
appreciable contribution to the radiation density at the BBN and CMB
epochs.

We further assume that the entire energy density transferred to the dark
sector by inflaton decays is rapidly converted, by the end of reheating,
into an internally thermalised bath of massless dark photons. More
specifically, the massive dark Higgs bosons and fermions initially
produced in the decay are assumed to decay or annihilate efficiently
into dark photons, leaving no sizeable symmetric massive relic
component. Consequently, the entire dark sector energy density
contributes to dark radiation, $\rho'_{\rm DR}=\rho'$. This assumption
defines the dark radiation benchmark considered below; the microscopic
decay and annihilation processes responsible for this conversion are
not modelled explicitly.

Since the dark radiation bath contains only the two transverse
polarisation states of the massless dark photon, its relativistic
particle content remains unchanged between reheating and the CMB epoch.
Therefore,
\begin{equation}
    \left[
        \frac{g_{*s}'(T'_{\rm RH})}
             {g_{*s}'(T'_{\rm CMB})}
    \right]^{1/3}
    = 1 .
    \label{eq:dark-entropy-ratio}
\end{equation}


\subsection{Constraints from \texorpdfstring{$\Delta N_{\rm eff}$}
{Delta Neff}}
\label{subsec:delta-neff}

The contribution of the dark radiation bath to the effective number of
relativistic neutrino species is
\begin{equation}
    \Delta N_{\rm eff}
    =
    \frac{4}{7}
    \left(
        \frac{11}{4}
    \right)^{4/3}
    g'_{*}(T'_{\rm CMB})
    \left(
        \frac{T'_{\rm CMB}}
             {T_{\rm CMB}}
    \right)^4.
    \label{eq:delta-neff-temperature}
\end{equation}
Within this effective-radiation description, and assuming adiabatic
evolution in each sector after reheating, the relation
$\rho'/\rho=\epsilon$, together with
$g_*(T_{\rm RH})\simeq g_{*s}(T_{\rm RH})$ and
$g'_*(T'_{\rm RH})\simeq g'_{*s}(T'_{\rm RH})$, gives
\begin{equation}
\Delta N_{\rm eff}
=
\frac{4}{7}
\left(\frac{11}{4}\right)^{4/3}
g'_\ast(T'_{\rm CMB})
\left[
\frac{g'_{\ast s}(T'_{\rm RH})}
     {g_{\ast s}(T_{\rm RH})}
\right]^{1/3}
\left[
\frac{g_{\ast s}(T_{\rm CMB})}
     {g'_{\ast s}(T'_{\rm CMB})}
\right]^{4/3}
\epsilon ,
\label{eq:DeltaNeff_epsilon}
\end{equation}
We take
\begin{equation}
g_\ast(T_{\rm CMB}) = 3.36,
\qquad
g_{\ast s}(T_{\rm CMB}) = 3.91,
\qquad
g_\ast(T_{\rm RH}) \simeq g_{\ast s}(T_{\rm RH}) = 95.25,
\end{equation}
where the latter value corresponds to
$T_{\rm RH}=100~{\rm GeV}$. Assuming in addition that
$g'_*(T'_{\rm CMB})\simeq g'_{*s}(T'_{\rm CMB})$,
the observational upper limit on the reheating asymmetry may then be written as
\begin{equation}
\epsilon_{\rm max}
=
0.337
\left[
\frac{g'_{\ast s}(T'_{\rm CMB})}
     {g'_{\ast s}(T'_{\rm RH})}
\right]^{1/3}
\Delta N_{\rm eff}.
\end{equation}
Using the Planck bound, the current combined BBN+CMB+BAO bound on a
positive contribution to $\Delta N_{\rm eff}$, and the projected
CMB-S4 sensitivity $\Delta N_{\rm eff}\simeq 0.06$ at $95\%$ C.L.,
adopted here as a representative next-generation CMB benchmark,
gives
\cite{Planck:2018vyg,Goldstein:2026iuu,Abazajian:2022CMB-S4}
\begin{align}
\epsilon^{\rm Planck}_{\rm max}
&\simeq
0.096
\left[
\frac{g'_{\ast s}(T'_{\rm CMB})}
     {g'_{\ast s}(T'_{\rm RH})}
\right]^{1/3}
\left(\frac{\Delta N_{\rm eff}}{0.284}\right),
\\
    \epsilon_{\rm max}^{\rm BBN+CMB+BAO}
  &\simeq
    0.0361
    \left[
    \frac{g'_{*s}(T'_{\rm CMB})}
         {g'_{*s}(T'_{\rm RH})}
    \right]^{1/3}
    \left(
    \frac{\Delta N_{\rm eff}}{0.107}
    \right). \label{epsACT}
\\
\epsilon^{\rm CMB-S4}_{\rm max}
&\simeq
0.0202
\left[
\frac{g'_{\ast s}(T'_{\rm CMB})}
     {g'_{\ast s}(T'_{\rm RH})}
\right]^{1/3}
\left(\frac{\Delta N_{\rm eff}}{0.06}\right).
\end{align}
For comparable relativistic degrees of freedom in the two sectors, the
current combined BBN+CMB+BAO bound allows a temperature ratio as large as
\begin{equation}
    \left(\frac{T'}{T}\right)_{\rm max}
    \sim 0.44 .
    \label{eq:temperature_ratio_bound}
\end{equation}
This is much larger than the temperature ratio predicted in the
4-body regime. To estimate the largest dark radiation contribution in this regime within the factorised treatment, we use the maximised 4-body rate derived in Sec.~\ref{subsec:4-body}. For illustration, we take
\begin{equation}
g'=g,\qquad
g'_\ast(T'_{\rm RH})
=
g'_{\ast s}(T'_{\rm RH})
=
g'_\ast(T'_{\rm CMB})
=
g'_{\ast s}(T'_{\rm CMB})
=2.
\end{equation}
For each value of
\begin{equation}
    z\equiv\frac{m_\phi}{M_{H'}},
\end{equation}
we maximise over the kinematically allowed range
\begin{equation}
    0<\mu<\frac14,
\end{equation}
where $\mu=m_{f'}/m_\phi$. The maximum occurs at an intermediate
value of $\mu$, rather than at the kinematic endpoint $\mu=1/4$,
where the 4-body phase space vanishes. Numerically, the optimum
depends only weakly on $z$ and approaches
\begin{equation}
    \mu_{\rm opt}\simeq0.125
\end{equation}
as $z\rightarrow1^-$. The corresponding dark radiation contribution
approaches
\begin{equation}
\Delta N_{\rm eff}^{(4),\max}
\simeq 3.3\times 10^{-11},
\qquad z\to 1^- .
\end{equation}
Neither present constraints nor the next-generation CMB sensitivity
benchmark considered here can therefore probe the 4-body regime through
$\Delta N_{\rm eff}$.

The situation changes significantly in the intermediate 3-body
regime,
\begin{equation}
    M_{H'}<m_\phi<2M_{H'}.
\end{equation}
As discussed in Sec.~\ref{sec:threebodyr}, the corresponding reheating asymmetry
depends on the dark Yukawa coupling. For each value of $z$, we
therefore maximise the physical 3-body decay rate over the
kinematically allowed range
\begin{equation}
    0<y'<\frac{z-1}{2},
\end{equation}
including both the final-state fermion mass and the finite width of
the intermediate dark Higgs.

For the same benchmark choice, the resulting 3-body contribution to
$\Delta N_{\rm eff}$ grows rapidly as the 2-body threshold is approached.
Very close to $z=2$, however, the separation between the three- and
2-body descriptions is no longer uniform, since the virtual dark Higgs
propagator approaches its mass shell. In this narrow threshold region, we evaluate the four-fermion final state including finite Breit-Wigner widths in both dark Higgs propagators. We approximate the squared matrix element by the sum of the squared contributions from the two possible fermion-antifermion pairings, neglecting the interference between them. In the narrow-width limits, this prescription reproduces the single-resonant 3-body result below threshold and the double-resonant 2-body result above threshold, providing a smooth interpolation across $z=2$. Using the permutation symmetry of the 4-body phase space, the two
squared pairing contributions yield equal integrated rates. Neglecting
their interference, the decay width used in the threshold region can therefore be written as
\begin{eqnarray}
    \Gamma'_{4,\mathrm{BW}}
    &=&
    \frac{g'^2 y'^4 v_\phi^2}
         {2048\pi^5 m_\phi^3} \label{eq:four_body_finite_width}
   \\
    &\times&
    \int_{4m_{f'}^2}^{(m_\phi-2m_{f'})^2}
    \!\!\!\!\!\!\!\!ds_1
    \int_{4m_{f'}^2}^{(m_\phi-\sqrt{s_1})^2}
     \!\!\!\!\!ds_2
    \frac{
        s_1 s_2\,
        \lambda^{1/2}(m_\phi^2,s_1,s_2)
    }{
        D_{H'}(s_1)D_{H'}(s_2)
    }
    \left(
        1-\frac{4m_{f'}^2}{s_1}
    \right)^{3/2}
    \left(
        1-\frac{4m_{f'}^2}{s_2}
    \right)^{3/2}.\nonumber
\end{eqnarray}
Here,
\begin{equation}
    D_{H'}(s)
    \equiv
    \left(s-M_{H'}^2\right)^2
    +M_{H'}^2\Gamma_{H'}^2,
    \label{eq:dark_higgs_breit_wigner}
\end{equation}
where $\Gamma_{H'}$ is the dark Higgs width defined in Sec.~\ref{sec:threebodyr}.
The factor of two included in the prefactor of
Eq.~\eqref{eq:four_body_finite_width} accounts for the two squared
fermion-antifermion pairing contributions.

Within this treatment, after maximising over the dark Yukawa coupling, we find
\begin{equation}
    \Delta N_{\rm eff}^{\max}
    \simeq 5.7\times10^{-2}
    \qquad \text{for} \qquad
    z=1.999, \label{eq:dneff_three_example}
\end{equation}
while exactly at the dark Higgs pair-production threshold,
\begin{equation}
    \Delta N_{\rm eff}^{\max}(z=2)
    \simeq 7.8\times10^{-2}.
    \label{eq:dneff_threshold}
\end{equation}
The Planck bound remains non-restrictive for the benchmark considered here.
The current combined BBN+CMB+BAO bound,
$\Delta N_{\rm eff}<0.107$, is reached at
\begin{equation}
   z \simeq 2.001 ,
    \label{eq:z_current_bound}
\end{equation}
whereas a sensitivity $\Delta N_{\rm eff}\simeq0.060$, corresponding to the
CMB-S4 forecast and used here as a representative next-generation CMB
benchmark, is reached at
\begin{equation}
    z \simeq 1.999 .
    \label{eq:z_cmbs4}
\end{equation}
Accordingly, the quoted crossing points should be regarded as benchmark
estimates rather than precision determinations. The finite-width treatment shows that the transition across the
dark Higgs pair-production threshold is smooth. On the 2-body side,
the current bound is satisfied only within a narrow interval immediately
above the threshold, whereas larger values of $z$ are excluded for this
benchmark. A future CMB experiment reaching a sensitivity of order
$\Delta N_{\rm eff}\simeq0.06$ would probe slightly below the threshold.

\begin{figure}[htbp!]
    \centering
\includegraphics[width=1.\textwidth]
    {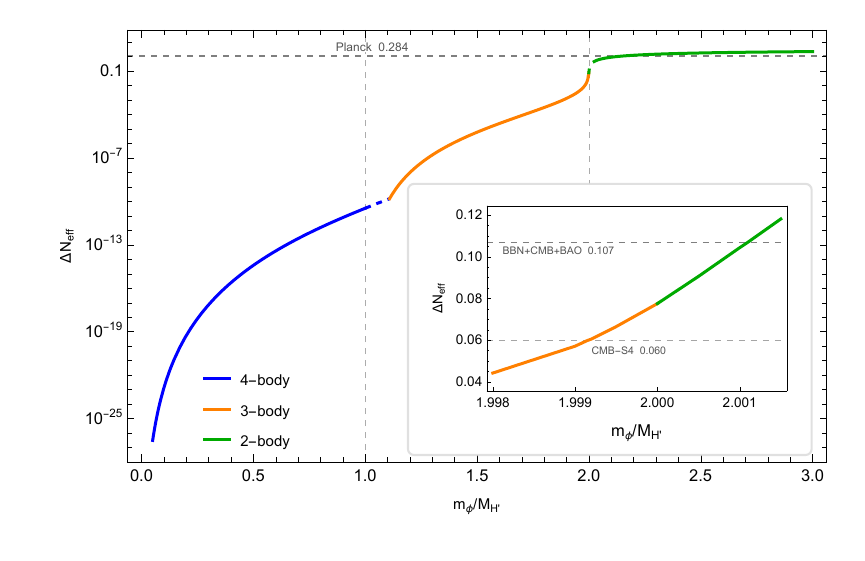}
    
    \caption{Maximum contribution to the effective number of relativistic neutrino species,
$\Delta N_{\rm eff}$, as a function of the inflaton-to-dark Higgs mass
ratio $z=m_\phi/M'_H$.
The blue solid curve shows the maximised 4-body contribution in the
regime $z<1$, where both dark Higgs bosons are necessarily off shell.
The blue dashed segment for $1<z\lesssim1.11$ shows the continuation
of the double-off-shell contribution, which remains dominant after
maximisation over the dark Yukawa coupling within the factorised
4-body treatment, even though the single-resonant 3-body
channel is already kinematically accessible.
The orange curve shows the maximised 3-body contribution once it
becomes dominant.
Near the pair-production threshold $z=2$, an approximate finite-width treatment of the four-fermion final state provides a smooth interpolation between the single-resonant 3-body and fully on-shell 2-body regimes.
The green curve shows the 2-body result above this threshold.
The vertical dashed lines at $z=1$ and $z=2$ indicate the corresponding
kinematic thresholds.
The horizontal dashed lines indicate the Planck bound, the current
combined BBN+CMB+BAO bound $\Delta N_{\rm eff}<0.107$ at $95\%$ C.L.,
and the CMB-S4 forecast $\Delta N_{\rm eff}\simeq0.060$, used here as a
representative benchmark for next-generation CMB sensitivity.
The inset enlarges the region around $z=2$, showing explicitly the
finite-width transition and the current and forecast sensitivities.
We take $g'=g$, a single dominant fermion species, and
$g'_\ast(T'_{\rm RH})=g'_{\ast s}(T'_{\rm RH})
=g'_\ast(T'_{\rm CMB})=g'_{\ast s}(T'_{\rm CMB})=2$.
     }
    \label{fig:neff_decay_regimes}
\end{figure}

The maximum contribution to $\Delta N_{\rm eff}$ across the three decay regimes is summarized in
Fig.~\ref{fig:neff_decay_regimes}. The 4-body contribution is strongly
suppressed for $m_\phi<M_{H'}$, while the 3-body contribution grows
rapidly as the dark Higgs pair-production threshold is approached. Above
$m_\phi=2M_{H'}$, the fully on-shell 2-body channel becomes kinematically
accessible.

\section{Conclusions}
\label{sec:conclusions}

In this work, we have studied the asymmetric post-inflationary reheating in a framework
containing two sequestered MSSM-like sectors with different SUSY-breaking
scales. The Affleck-Dine mechanism operates independently in the visible and dark
sectors, generating the corresponding matter-antimatter asymmetries. Together with the
different confinement scales, this provides a possible explanation for the comparable
present-day baryonic and dark matter abundances without requiring the particles in the
two sectors to have comparable masses.

The hierarchy between the inflaton and dark Higgs masses plays a central role in
determining how efficiently the dark sector is reheated. While the inflaton can decay
efficiently into lighter visible-sector states, its decay into the dark sector can
involve zero, one, or two off-shell dark Higgs propagators, with the relevant topology
determined primarily by the inflaton-dark Higgs mass hierarchy and, in the intermediate
region, also by the dark fermion mass. The resulting suppression of the dark sector
decay width directly translates into a smaller dark radiation energy density and,
consequently, into a lower dark sector temperature.

When the inflaton is lighter than the dark Higgs, both dark Higgs bosons are virtual and
the leading decay into the dark sector is a 4-body process. We have refined the
treatment of this channel by retaining the dark fermion mass, the full momentum
dependence of the two virtual propagators, and the full integration over the
kinematically allowed 4-body phase space within the factorised treatment adopted
here. We find that the maximal decay rate occurs for an intermediate value of the dark
Yukawa coupling rather than at its kinematic upper limit, where the available phase
space vanishes. In the strongly hierarchical limit, the characteristic twelfth-power
suppression with the inflaton-to-dark Higgs mass ratio is recovered.

For the dark radiation analysis, we have assumed that the massive particles initially
produced by inflaton decays efficiently decay or annihilate into an internally
thermalized bath of massless dark photons, so that the entire injected dark sector
energy density ultimately redshifts as radiation. The 4-body channel therefore leads
to an extremely cold dark sector. For the benchmark considered here, its maximal
contribution to dark radiation approaches only $\Delta N_{\rm eff} \simeq 3.3\times10^{-11}$
near the boundary of the 4-body regime. This is many orders of magnitude below both
present constraints and the sensitivities of the future CMB measurements considered
here. The strong suppression of the dark sector temperature also prevents a deconfined
dark-QCD phase over most of the strongly hierarchical parameter region studied and
strongly suppresses dark weak thermalisation for the representative benchmarks
considered.

The situation changes substantially in the intermediate region
$M_{H'} < m_\phi < 2M_{H'}$, where the single-resonant 3-body channel can become
kinematically accessible, depending on the dark fermion mass. Immediately above
$m_\phi = M_{H'}$, however, the kinematic restriction on the dark Yukawa coupling implies
that the maximised double-off-shell four-fermion contribution can still dominate; for the
benchmark considered here, and within the factorised 4-body treatment adopted in this
work, the maximised 3-body contribution takes over at $m_\phi/M_{H'} \simeq 1.11$.
This crossover occurs at $\Delta N_{\rm eff} \sim 10^{-10}$ and is therefore irrelevant
for dark radiation searches. At larger mass ratios, the 3-body contribution grows
rapidly as the dark Higgs pair-production threshold is approached. Very close to
$m_\phi = 2M_{H'}$, the pure 3-body description is no longer uniform, and we use an
approximate finite-width treatment of the four-fermion final state to interpolate
smoothly between the single- and double-resonant regions. For the benchmark considered
here, this gives $\Delta N_{\rm eff}^{\rm max} \simeq 7.8\times10^{-2}$ at
$m_\phi = 2M_{H'}$.

Once the inflaton is heavy enough to produce two dark Higgs bosons on shell, the
suppression associated with virtual propagators disappears. For comparable inflaton
couplings to the two sectors, reheating then becomes much more symmetric away from
threshold, so a sizeable temperature hierarchy requires the inflaton mass to remain close
to the dark Higgs pair-production threshold.

Overall, the different kinematic regions span the evolution from an extremely weakly
reheated dark sector when both dark Higgs bosons are virtual, through the opening and
subsequent dominance of the single-resonant 3-body channel, to the fully on-shell
2-body regime. The strongly hierarchical 4-body region is effectively inaccessible
to dark radiation searches, whereas the vicinity of the dark Higgs pair-production
threshold can give rise to an observable contribution to $\Delta N_{\rm eff}$. For this
reason, the current combined BBN+CMB+BAO bound already excludes the 2-body regime
studied here except for a narrow window immediately above threshold
($m_\phi/M_{H'} \simeq 2.001$), and it is this same window - extending slightly further,
to $m_\phi/M_{H'} \simeq 1.999$, under a next-generation CMB sensitivity of order
$\Delta N_{\rm eff} \simeq 0.06$ - that precision dark radiation measurements are best
positioned to probe. Together with the finite-width treatment showing that the transition
across the threshold is smooth, this identifies the dark Higgs pair-production threshold
as the most promising place to look for the imprint of asymmetric reheating in the
dark radiation content of the Universe.
 
Our analysis focused exclusively on perturbative decays of the inflaton field. While these generically dominate at the end of the reheating process, there could be an initial period of non-perturbative decay through parametric resonance, known as preheating. During this period, particles much heavier than the inflaton field may be produced on-shell, particularly in the broad resonance regime with $q=2gv_\phi\Phi/m_\phi^2\gg 1$, where $\Phi$ denotes the inflaton's oscillation amplitude \cite{Kofman:1994rk, Kofman:1997yn}. While this could potentially lead to an efficient production of dark Higgs scalars outside the 2-body decay regime, we note that we are mostly interested in scenarios with low reheating temperatures, where the Affleck-Dine mechanism yields the observed baryon asymmetry. This implies a relatively long reheating process, with small couplings such that preheating should occur mostly in the narrow resonance regime, where heavy particle production is heavily suppressed. Our analysis is therefore applicable in this regime, although it would be interesting to further explore, in the future, the role of non-perturbative effects in asymmetric reheating scenarios.

While we have assumed that the Higgs is the lightest scalar state in both sectors, and therefore the main decay channel, a similar analysis would apply if one of the squark and slepton fields is lighter than the Higgs. In this case the relation between the fermion and scalar masses would be different from the one considered in this work, although this may be taken into account in the value of the effective Yukawa coupling. 

Our results are also applicable to non-SUSY scenarios with a Standard Model-like dark sector, since in this case only the Higgs fields in both sectors may couple directly to the inflaton field, assuming that the latter is a singlet. This therefore broadens the scope of our analysis, although in such scenarios the hierarchy between the dark QCD and electroweak scales may be different, and the dark matter-baryon coincidence problem must be addressed in an alternative way.


\acknowledgments

CC is supported by the FCT - Funda\c{c}\~{a}o para a Ci\^{e}ncia e Tecnologia, I.P. project Grant No. IN1234CEECINST/00099/2021. This work was supported by national funds by FCT, through the research project with DOI identifier 10.54499/UID/04564/2025 , and by the project 10.54499/2024.00252.CERN funded by measure RE-C06-i06.m02 – ``Reinforcement of funding for International Partnerships in Science, Technology and Innovation'' of the Recovery and Resilience Plan - RRP, within the framework of the financing contract signed between the Recover Portugal Mission Structure (EMRP) and the Foundation for Science and Technology I.P. (FCT), as an intermediate beneficiary.


\appendix 

\section{Analytical solution of the Boltzmann equations}
\label{app:boltzmann}

In this appendix, we derive the analytical solutions of the Boltzmann
equations describing the asymmetric reheating of the visible and dark
sectors, introduced in Section~\ref{sec:reheating}:
\begin{align}
    \dot{\rho}_{\phi}
    +3H\rho_{\phi}
    &=
    -\Gamma_{\phi}\rho_{\phi},
    \label{eq:app-boltzmann-inflaton}
    \\
    \dot{\rho}
    +4H\rho
    &=
    \Gamma\rho_{\phi},
    \label{eq:app-boltzmann-visible}
    \\
    \dot{\rho}'
    +4H\rho'
    &=
    \Gamma'\rho_{\phi}.
    \label{eq:app-boltzmann-dark}
\end{align}
Here,
\begin{equation}
    \Gamma_{\phi}
    =
    \Gamma+\Gamma'
\end{equation}
is the total inflaton decay width, while $\Gamma$ and $\Gamma'$ are the
partial widths into the visible and dark sectors, respectively. The
Hubble parameter is
\begin{equation}
    H
    =
    \left(
        \frac{
            \rho_{\phi}+\rho+\rho'
        }{
            3M_{\rm Pl}^{2}
        }
    \right)^{1/2}.
    \label{eq:app-hubble}
\end{equation}

At the end of inflation, $t=t_{\rm e}$, we impose
\begin{equation}
    \rho_{\phi}(t_{\rm e})
    =
    V(\phi_{\rm e}),
    \qquad
    \rho(t_{\rm e})
    =
    \rho'(t_{\rm e})
    =
    0.
    \label{eq:app-initial-conditions}
\end{equation}
The end time of inflation is approximately
\begin{equation}
    t_{\rm e}
    \simeq
    \left[
        \frac{
            4M_{\rm Pl}^{2}
        }{
            3V(\phi_{\rm e})
        }
    \right]^{1/2}.
    \label{eq:app-end-time}
\end{equation}

The evolution can be divided into an early inflaton-dominated epoch
and a later radiation-dominated epoch.

\subsection{Inflaton-dominated epoch}
\label{app:inflaton-domination}

While the inflaton dominates the total energy density, the Universe
behaves as matter dominated,
\begin{equation}
    H(t)
    \simeq
    \frac{2}{3t}.
    \label{eq:app-hubble-matter}
\end{equation}
The inflaton equation can then be integrated directly,
\begin{equation}
    \rho_{\phi}(t)
    =
    \rho_{\phi}(t_{\rm e})
    \exp\left[
        -\Gamma_{\phi}
        \left(
            t-t_{\rm e}
        \right)
    \right]
    \left(
        \frac{t_{\rm e}}{t}
    \right)^{2}.
    \label{eq:app-rho-phi-general-matter}
\end{equation}
In the absence of significant inflaton depletion, the matter-dominated
Friedmann relation gives
\begin{equation}
    \rho_\phi(t) \simeq \frac{4M_{\rm Pl}^2}{3t^2}. \label{eq:reduced_rho}
\end{equation}
Using $\rho_\phi(t_e)=V(\phi_e)\simeq
4M_{\rm Pl}^2/(3t_e^2)$ in Eq.~(\ref{eq:app-rho-phi-general-matter}), the inflaton density may more
generally be written as
\begin{equation}
    \rho_{\phi}(t)
    \simeq
    \frac{
        4M_{\rm Pl}^{2}
    }{
        3t^{2}
    }
    \exp\left[
        -\Gamma_{\phi}t
        \left(
            1-\frac{t_{\rm e}}{t}
        \right)
    \right].
    \label{eq:app-rho-phi-matter}
\end{equation}
For the leading early-time solution of the radiation equations,
$\Gamma_\phi t\ll 1$, we neglect the exponential depletion in the source
term and use Eq.~(\ref{eq:reduced_rho}).

For the visible sector, Eq.~\eqref{eq:app-boltzmann-visible} reduces to
\begin{equation}
    \dot{\rho}
    +
    \frac{8}{3t}\rho
    =
    \Gamma
    \frac{
        4M_{\rm Pl}^{2}
    }{
        3t^{2}
    }.
    \label{eq:app-visible-matter-equation}
\end{equation}
Multiplying by the integrating factor $t^{8/3}$ gives
\begin{equation}
    \frac{d}{dt}
    \left(
        t^{8/3}\rho
    \right)
    =
    t^{8/3}
    \Gamma
    \frac{
        4M_{\rm Pl}^{2}
    }{
        3t^{2}
    }.
\end{equation}
After integration and using $\rho(t_{\rm e})=0$, one obtains
\begin{equation}
    \rho(t)
    =
    \frac{4M_{\rm Pl}^2}{5t}\,
    \Gamma
    \left[
        1-\left(\frac{t_e}{t}\right)^{5/3}
    \right].
    \label{eq:app-rho-visible-matter}
\end{equation}
The dark sector equation has the same form, yielding
\begin{equation}
    \rho'(t)
    =
    \frac{4M_{\rm Pl}^2}{5t}\,
    \Gamma'
    \left[
        1-\left(\frac{t_e}{t}\right)^{5/3}
    \right].
    \label{eq:app-rho-dark-matter}
\end{equation}

\subsection{Radiation-dominated epoch}
\label{app:radiation-domination}

After most of the inflaton abundance has decayed, the Universe becomes
radiation dominated and
\begin{equation}
    H(t)
    =
    \frac{1}{2t}.
    \label{eq:app-hubble-radiation}
\end{equation}
We approximate the transition as occurring at the inflaton lifetime,
\begin{equation}
    \tau_{\phi}
    \equiv
    \Gamma_{\phi}^{-1}.
    \label{eq:app-inflaton-lifetime}
\end{equation}
The inflaton density is then
\begin{align}
    \rho_{\phi}(t)
    &=
    \rho_{\phi}(\tau_{\phi})
    \exp\left[
        -\Gamma_{\phi}
        \left(
            t-\tau_{\phi}
        \right)
    \right]
    \left(
        \frac{\tau_{\phi}}{t}
    \right)^{3/2}
    \nonumber
    \\
    &\simeq
    \frac{
        4M_{\rm Pl}^{2}
    }{
        3\tau_{\phi}^{2}
    }
    \exp\left[
        -\Gamma_{\phi}t
        \left(
            1-\frac{t_{\rm e}}{t}
        \right)
    \right]
    \left(
        \frac{\tau_{\phi}}{t}
    \right)^{3/2}.
    \label{eq:app-rho-phi-radiation}
\end{align}
For a simple late-time analytical approximation, we neglect the
residual energy injection from inflaton decays once
$t\sim\tau_\phi$ and let the radiation densities subsequently
redshift as $a^{-4}\propto t^{-2}$. We normalise these late-time
solutions by extrapolating the early-time expressions to
$t=\tau_\phi$. Since $\Gamma_\phi\tau_\phi=1$, this matching should
be understood as an approximate interpolation between the two
asymptotic regimes rather than as a controlled expansion at
$t=\tau_\phi$. This gives
\begin{equation}
    \rho(t)
    \simeq
    \frac{4}{5}
    \frac{
        M_{\rm Pl}^{2}
    }{
        \tau_{\phi}
    }
    \Gamma
    \left[
        1-
        \left(
            \frac{t_{\rm e}}
                 {\tau_{\phi}}
        \right)^{5/3}
    \right]
    \left(
        \frac{\tau_{\phi}}{t}
    \right)^{2},
    \label{eq:app-rho-visible-radiation}
\end{equation}
and
\begin{equation}
    \rho'(t)
    \simeq
    \frac{4}{5}
    \frac{
        M_{\rm Pl}^{2}
    }{
        \tau_{\phi}
    }
    \Gamma'
    \left[
        1-
        \left(
            \frac{t_{\rm e}}
                 {\tau_{\phi}}
        \right)^{5/3}
    \right]
    \left(
        \frac{\tau_{\phi}}{t}
    \right)^{2}.
    \label{eq:app-rho-dark radiation}
\end{equation}

The ratio of the two radiation densities is therefore fixed by the
corresponding partial decay widths in both regimes,
\begin{equation}
    \frac{\rho'(t)}{\rho(t)}
    =
    \frac{\Gamma'}{\Gamma}
    \equiv
    \epsilon.
    \label{eq:app-density-ratio}
\end{equation}
Thus, the reheating asymmetry is entirely determined by the relative
branching of the inflaton into the two sectors.

\section{Inflaton 2-body decay}
\label{app:2-body}

We consider the decay of the inflaton into two identical scalar
particles,
\begin{equation}
    \phi(p)
    \longrightarrow
    H(k_1)+H(k_2),
\end{equation}
with
\begin{equation}
    p=k_1+k_2,
    \qquad
    p^2=m_\phi^2,
    \qquad
    k_1^2=k_2^2=M_H^2.
\end{equation}
The relevant interaction is
\begin{equation}
    \mathcal{L}_{\rm int}
    \supset
    -\frac{1}{2}g v_\phi\,\phi H^2,
    \label{eq:app-2-body-interaction}
\end{equation}
which gives the matrix element
\begin{equation}
    i\mathcal{M}
    =
    -igv_\phi.
\end{equation}
Therefore,
\begin{equation}
    |\mathcal{M}|^2
    =
    g^2v_\phi^2.
    \label{eq:app-2-body-matrix-element}
\end{equation}

The differential decay width is
\begin{equation}
    d\Gamma
    =
    \frac{1}{2m_\phi}
    |\mathcal{M}|^2
    d\Phi_2,
    \label{eq:app-2-body-general-width}
\end{equation}
where the 2-body phase-space element is
\begin{equation}
    d\Phi_2
    =
    (2\pi)^4
    \delta^{(4)}
    \left(
        p-k_1-k_2
    \right)
    \frac{d^3\mathbf{k}_1}
         {(2\pi)^3 2E_1}
    \frac{d^3\mathbf{k}_2}
         {(2\pi)^3 2E_2}.
    \label{eq:app-2-body-phase-space}
\end{equation}

In the inflaton rest frame,
\begin{equation}
    p^\mu
    =
    \left(
        m_\phi,\mathbf{0}
    \right),
\end{equation}
and the two final-state particles have equal and opposite
three-momenta,
\begin{equation}
    \mathbf{k}_1
    =
    -\mathbf{k}_2.
\end{equation}
Their magnitude is
\begin{equation}
    |\mathbf{k}|
    =
    \frac{1}{2m_\phi}
    \lambda^{1/2}
    \left(
        m_\phi^2,M_H^2,M_H^2
    \right),
    \label{eq:app-2-body-momentum-kallen}
\end{equation}
where
\begin{equation}
    \lambda(x,y,z)
    =
    x^2+y^2+z^2-2xy-2xz-2yz
    \label{eq:app-kallen-function}
\end{equation}
is the Källén function. For identical final-state masses,
\begin{equation}
    |\mathbf{k}|
    =
    \frac{m_\phi}{2}
    \left(
        1-\frac{4M_H^2}{m_\phi^2}
    \right)^{1/2}.
    \label{eq:app-2-body-momentum}
\end{equation}

After integrating over the solid angle, the 2-body phase space is
\begin{equation}
    \int d\Phi_2
    =
    \frac{1}{8\pi}
    \frac{
        \lambda^{1/2}
        \left(
            m_\phi^2,M_H^2,M_H^2
        \right)
    }{
        m_\phi^2
    }
    =
    \frac{1}{8\pi}
    \left(
        1-\frac{4M_H^2}{m_\phi^2}
    \right)^{1/2}.
    \label{eq:app-2-body-integrated-phase-space}
\end{equation}

Since the final-state particles are identical, a symmetry factor
$1/2!$ must be included. The total decay width is therefore
\begin{align}
    \Gamma_{\phi\rightarrow HH}
    &=
    \frac{1}{2!}
    \frac{1}{2m_\phi}
    |\mathcal{M}|^2
    \int d\Phi_2
    \nonumber
    \\
    &=
    \frac{
        g^2v_\phi^2
    }{
        32\pi m_\phi
    }
    \left(
        1-\frac{4M_H^2}{m_\phi^2}
    \right)^{1/2}.
    \label{eq:app-2-body-width}
\end{align}
The decay is kinematically allowed for
\begin{equation}
    m_\phi>2M_H.
\end{equation}
In the limit $M_H\ll m_\phi$, the width reduces to
\begin{equation}
    \Gamma_{\phi\rightarrow HH}
    \simeq
    \frac{
        g^2v_\phi^2
    }{
        32\pi m_\phi
    }.
    \label{eq:app-2-body-width-light-mass}
\end{equation}

The corresponding expression for the dark sector is obtained through
the replacements
\begin{equation}
    g\rightarrow g',
    \qquad
    M_H\rightarrow M_{H'}.
\end{equation}


\section{Inflaton 3-body decay}
\label{app:threebody}

In this appendix, we derive the 3-body decay width for the process
\begin{equation}
    \phi(p)
    \longrightarrow H'(k)+H'^{*}(q)
    \longrightarrow H'(k)+f'(p_1)+\bar f'(p_2),
    \label{eq:app_threebody_process}
\end{equation}
in the intermediate mass regime
\begin{equation}
    M_{H'}<m_\phi<2M_{H'}.
\end{equation}
We define
\begin{equation}
    q=p_1+p_2,
    \qquad
    s\equiv q^2.
\end{equation}
In contrast to the massless approximation, we retain the final-state
fermion mass throughout the calculation and use the benchmark relation
adopted in the main text,
\begin{equation}
    m_{f'} = y'M_{H'}.
\end{equation}
This is particularly important when the decay rate is maximised with respect
to the dark Yukawa coupling $y'$.

\subsection{Matrix element}

The relevant interaction terms are
\begin{equation}
    \mathcal{L}_{\rm int}
    \supset
    -\frac{1}{2}g'v_\phi\,\phi H'^2
    -y'H'\bar f'f'.
    \label{eq:app_threebody_Lint}
\end{equation}
The tree-level matrix element is therefore
\begin{equation}
    i\mathcal{M}_3
    =
    (-ig'v_\phi)
    \frac{i}
    {s-M_{H'}^2+iM_{H'}\Gamma_{H'}}
    (-iy')
    \bar u(p_1)v(p_2),
    \label{eq:app_threebody_M}
\end{equation}
where the finite width of the virtual dark Higgs has been retained. This is
required near the upper boundary $m_\phi\rightarrow2M_{H'}^{-}$, where the
virtual state approaches its mass shell.

Summing over the final-state fermion spins gives
\begin{align}
    \sum_{\rm spins}
    \left|
    \bar u(p_1)v(p_2)
    \right|^2
    &=
    {\rm Tr}
    \left[
    (\slashed p_1+m_{f'})
    (\slashed p_2-m_{f'})
    \right]
    \nonumber\\
    &=
    4\left(p_1\cdot p_2-m_{f'}^2\right).
\end{align}
Using
\begin{equation}
    s=(p_1+p_2)^2
    =
    2m_{f'}^2+2p_1\cdot p_2,
\end{equation}
one obtains
\begin{equation}
    \sum_{\rm spins}
    \left|
    \bar u(p_1)v(p_2)
    \right|^2
    =
    2\left(s-4m_{f'}^2\right).
    \label{eq:app_threebody_spin}
\end{equation}
The spin-summed squared matrix element is therefore
\begin{equation}
    \overline{|\mathcal{M}_3|^2}
    =
    \frac{
    2g'^2y'^2v_\phi^2
    \left(s-4m_{f'}^2\right)
    }
    {
    (s-M_{H'}^2)^2
    +M_{H'}^2\Gamma_{H'}^2
    }.
    \label{eq:app_threebody_M2}
\end{equation}
No additional symmetry factor is required, since the three final-state
particles $H'$, $f'$, and $\bar f'$ are distinct.

\subsection{3-body phase space}

The 3-body phase-space measure can be factorised as
\begin{equation}
    d\Phi_3(p;k,p_1,p_2)
    =
    \frac{ds}{2\pi}\,
    d\Phi_2(p;k,q)\,
    d\Phi_2(q;p_1,p_2).
    \label{eq:app_threebody_phasefactor}
\end{equation}
For the first 2-body phase-space factor,
corresponding to $\phi(p)\rightarrow H'(k)+H'^{*}(q)$,
\begin{equation}
    d\Phi_2(p;k,q)
    =
    \frac{1}{16\pi^2}
    \frac{|\mathbf{k}|}{m_\phi}\,
    d\Omega,
\end{equation}
where, in the inflaton rest frame,
\begin{equation}
    |\mathbf{k}|
    =
    \frac{
    \lambda^{1/2}(m_\phi^2,M_{H'}^2,s)
    }
    {2m_\phi}.
\end{equation}
Hence
\begin{equation}
    d\Phi_2(p;k,q)
    =
    \frac{
    \lambda^{1/2}(m_\phi^2,M_{H'}^2,s)
    }
    {32\pi^2m_\phi^2}
    d\Omega.
    \label{eq:app_threebody_phi2first}
\end{equation}

For the decay of the virtual Higgs,
$H'^{*}(q)\rightarrow f'(p_1)+\bar f'(p_2)$,
we work in the rest frame of $q$. The fermion momentum is
\begin{equation}
    |\mathbf{p}_1^*|
    =
    \frac{\sqrt{s}}{2}
    \left(
    1-\frac{4m_{f'}^2}{s}
    \right)^{1/2},
\end{equation}
and therefore
\begin{equation}
    d\Phi_2(q;p_1,p_2)
    =
    \frac{1}{32\pi^2}
    \left(
    1-\frac{4m_{f'}^2}{s}
    \right)^{1/2}
    d\Omega^*.
    \label{eq:app_threebody_phi2second}
\end{equation}

Combining the two factors and integrating over the solid angles,
\begin{equation}
    \int d\Omega=4\pi,
    \qquad
    \int d\Omega^*=4\pi,
\end{equation}
gives
\begin{equation}
    \frac{d\Phi_3}{ds}
    =
    \frac{
    \lambda^{1/2}(m_\phi^2,M_{H'}^2,s)
    }
    {128\pi^3m_\phi^2}
    \left(
    1-\frac{4m_{f'}^2}{s}
    \right)^{1/2}.
    \label{eq:app_threebody_dphi3}
\end{equation}

The invariant mass of the fermion pair is bounded from below by
\begin{equation}
    s_{\rm min}=4m_{f'}^2,
\end{equation}
while energy conservation requires
\begin{equation}
    M_{H'}+\sqrt{s}\leq m_\phi,
\end{equation}
and therefore
\begin{equation}
    s_{\rm max}
    =
    (m_\phi-M_{H'})^2.
    \label{eq:app_threebody_smax}
\end{equation}

\subsection{Differential and total decay widths}

The differential decay width is
\begin{equation}
    d\Gamma'_3
    =
    \frac{1}{2m_\phi}
    \overline{|\mathcal{M}_3|^2}
    d\Phi_3.
\end{equation}
Using Eqs.~\eqref{eq:app_threebody_M2} and
\eqref{eq:app_threebody_dphi3}, one obtains
\begin{equation}
\frac{d\Gamma'_3}{ds}
=
\frac{g'^2y'^2v_\phi^2}
     {128\pi^3m_\phi^3}
\,
\frac{
s\,\lambda^{1/2}(m_\phi^2,M_{H'}^2,s)
}{
(s-M_{H'}^2)^2+M_{H'}^2\Gamma_{H'}^2
}
\left(
1-\frac{4m_{f'}^2}{s}
\right)^{3/2}.
\label{eq:app_threebody_dGamma}
\end{equation}
The factor with power $3/2$ results from the combination of the
mass-dependent spin sum and the 2-body phase space of the fermion pair.

The total width is consequently
\begin{equation}
\Gamma'_3
=
\frac{g'^2y'^2v_\phi^2}
     {128\pi^3m_\phi^3}
\int_{4m_{f'}^2}^{(m_\phi-M_{H'})^2}
ds\,
\frac{
s\,\lambda^{1/2}(m_\phi^2,M_{H'}^2,s)
}{
(s-M_{H'}^2)^2+M_{H'}^2\Gamma_{H'}^2
}
\left(
1-\frac{4m_{f'}^2}{s}
\right)^{3/2}.
\label{eq:app_threebody_Gammafull}
\end{equation}

Introducing
\begin{equation}
    r\equiv\frac{M_{H'}}{m_\phi},
    \qquad
    x\equiv\frac{s}{m_\phi^2},
    \qquad
    \gamma\equiv\frac{\Gamma_{H'}}{M_{H'}},
\end{equation}
and using $m_{f'} = y'M_{H'}$, this becomes
\begin{equation}
    \Gamma'_3
    =
    \frac{g'^2y'^2v_\phi^2}
         {128\pi^3m_\phi}
    F_3(r,\gamma,y'),
\end{equation}
where
\begin{equation}
\begin{split}
F_3(r,\gamma,y')
={}&
\int_{4y'^2r^2}^{(1-r)^2}
dx\,
\frac{
x\,\lambda^{1/2}(1,r^2,x)
}{
(x-r^2)^2+r^4\gamma^2
}
\\
&\times
\left(
1-\frac{4y'^2r^2}{x}
\right)^{3/2}.
\end{split}
\label{eq:app_threebody_F3}
\end{equation}

The requirement that the 3-body final state be kinematically accessible,
\begin{equation}
    m_\phi\geq M_{H'}+2m_{f'},
\end{equation}
implies
\begin{equation}
    0<y'<\frac{z-1}{2},
    \qquad
    z\equiv\frac{m_\phi}{M_{H'}},
    \qquad
    1<z<2.
    \label{eq:app_threebody_ybound}
\end{equation}

If a single fermion species dominates the total width of the
intermediate dark Higgs, we use
\begin{equation}
 \Gamma_{H'}
\simeq
\Gamma_{H'\to f'\bar f'}
=
\frac{y'^2 M_{H'}}{8\pi}
\left(
    1-\frac{4m_{f'}^2}{M_{H'}^2}
\right)^{3/2}.
\end{equation}
Using $m_{f'} = y'M_{H'}$ therefore gives
\begin{equation}
    \gamma(y')
    =
    \frac{y'^2}{8\pi}
    \left(1-4y'^2\right)^{3/2}.
    \label{eq:app_threebody_gamma}
\end{equation}
Notice that the kinematic bound in
Eq.~\eqref{eq:app_threebody_ybound} automatically implies $y'<1/2$ in the
3-body regime, so that the decay $H'\rightarrow f'\bar f'$ is
kinematically allowed.

Away from the upper 2-body threshold and in the limit
$m_{f'}\rightarrow0$, the massive phase-space factor approaches unity and
the lower integration limit approaches zero, reproducing the massless
zero-width expression used as an approximation in the simpler limit.
Close to $m_\phi=2M_{H'}$, however, the finite-width treatment is essential.


\section{Inflaton 4-body decay}
\label{app:4-body}

In this appendix, we derive the factorised 4-body contribution
used in Sec.~\ref{subsec:4-body} for inflaton decay into four dark fermions mediated
by two off-shell dark Higgs bosons,

\begin{equation}
    \phi(p)
    \longrightarrow
    H'^{*}(q_1)+H'^{*}(q_2)
    \longrightarrow
    f'(p_1)+\bar f'(p_2)+f'(p_3)+\bar f'(p_4),
\end{equation}
where
\begin{equation}
    q_1=p_1+p_2,
    \qquad
    q_2=p_3+p_4,
\end{equation}
and
\begin{equation}
    s_1\equiv q_1^2,
    \qquad
    s_2\equiv q_2^2.
\end{equation}
We first consider the hierarchy
\begin{equation}
    m_\phi < M'_H ,
\end{equation}
for which both dark Higgs propagators are necessarily off shell.
The same factorised 4-body expression also applies in the
intermediate region $M'_H<m_\phi<2M'_H$ whenever the
single-resonant configuration is kinematically inaccessible,
\begin{equation}
    m_\phi < M'_H + 2m'_f .
\end{equation}
In this case, both fermion-pair invariant masses remain below the
dark Higgs pole, so that both propagators are again off shell.

We retain the final-state fermion mass throughout the calculation and
use the benchmark relation
\begin{equation}
    m_{f'} = y'M_{H'} .
\end{equation}
We also keep the full momentum dependence of both virtual scalar
propagators.
We follow the factorised treatment adopted in the main
text and include the statistical factor
\begin{equation}
    S=\frac{1}{2!2!}=\frac14
\end{equation}
for the two identical fermions and two identical antifermions.

\subsection{Matrix element}

The relevant interaction terms are
\begin{equation}
    \mathcal L_{\rm int}
    \supset
    -\frac12 g'v_\phi\,\phi H'^2
    -y'H'\bar f'f'.
\end{equation}
For the momentum assignment introduced above, the factorised
contribution associated with the pairing
$(p_1,p_2)(p_3,p_4)$ is
\begin{equation}
\begin{split}
    i\mathcal M_4
    ={}&
    (-ig'v_\phi)
    \frac{i}{s_1-M_{H'}^2}
    \frac{i}{s_2-M_{H'}^2}
    (-iy')^2
    \\
    &\times
    \left[\bar u(p_1)v(p_2)\right]
    \left[\bar u(p_3)v(p_4)\right].
\end{split}
\end{equation}
Since the physical phase space satisfies
\begin{equation}
    s_i \leq (m_\phi-2m'_f)^2 < M_H'^2 ,
\end{equation}
where the second inequality follows from
$m_\phi<M'_H+2m'_f$, the poles of the two scalar propagators
lie outside the kinematically accessible region. A finite-width
regulator is therefore not required in this double-off-shell regime.

For the first fermion-antifermion pair, the spin sum gives
\begin{align}
    \sum_{\rm spins}
    \left|\bar u(p_1)v(p_2)\right|^2
    &=
    {\rm Tr}\left[
    (\slashed p_1+m_{f'})
    (\slashed p_2-m_{f'})
    \right]
    \nonumber\\
    &=
    4\left(p_1\cdot p_2-m_{f'}^2\right).
\end{align}
Using
\begin{equation}
    s_1=(p_1+p_2)^2
       =2m_{f'}^2+2p_1\cdot p_2,
\end{equation}
we obtain
\begin{equation}
    \sum_{\rm spins}
    \left|\bar u(p_1)v(p_2)\right|^2
    =
    2\left(s_1-4m_{f'}^2\right).
\end{equation}
Analogously,
\begin{equation}
    \sum_{\rm spins}
    \left|\bar u(p_3)v(p_4)\right|^2
    =
    2\left(s_2-4m_{f'}^2\right).
\end{equation}
The spin-summed squared matrix element is therefore
\begin{equation}
    \sum_{\rm spins}|\mathcal M_4|^2
    =
    4g'^2y'^4v_\phi^2
    \frac{
    (s_1-4m_{f'}^2)(s_2-4m_{f'}^2)
    }{
    (s_1-M_{H'}^2)^2
    (s_2-M_{H'}^2)^2.
}
\end{equation}

\subsection{4-body phase space}

The 4-body phase-space element is
\begin{equation}
    d\Phi_4
    =
    (2\pi)^4
    \delta^{(4)}
    \left(
    p-\sum_{i=1}^{4}p_i
    \right)
    \prod_{i=1}^{4}
    \frac{d^3p_i}{(2\pi)^3\,2E_i}.
\end{equation}
Introducing the invariant masses $s_1$ and $s_2$, it can be factorised
as
\begin{equation}
    d\Phi_4
    =
    \frac{ds_1}{2\pi}
    \frac{ds_2}{2\pi}
    d\Phi_2(p;q_1,q_2)
    d\Phi_2(q_1;p_1,p_2)
    d\Phi_2(q_2;p_3,p_4).
\end{equation}
After integration over the angular variables,
\begin{equation}
    \int d\Phi_2(p;q_1,q_2)
    =
    \frac{1}{8\pi}
    \frac{
    \lambda^{1/2}(m_\phi^2,s_1,s_2)
    }{m_\phi^2},
\end{equation}
where
\begin{equation}
    \lambda(a,b,c)
    =
    a^2+b^2+c^2-2ab-2ac-2bc
\end{equation}
is the K\"all\'en function.

For each fermion-antifermion pair,
\begin{equation}
    \int d\Phi_2(q_i;f'\bar f')
    =
    \frac{1}{8\pi}
    \left(
    1-\frac{4m_{f'}^2}{s_i}
    \right)^{1/2}.
\end{equation}
The factorised 4-body phase space is consequently
\begin{equation}
    d\Phi_4
    =
    \frac{ds_1\,ds_2}{2048\pi^5}
    \frac{
    \lambda^{1/2}(m_\phi^2,s_1,s_2)
    }{m_\phi^2}
    \prod_{i=1}^{2}
    \left(
    1-\frac{4m_{f'}^2}{s_i}
    \right)^{1/2}.
\end{equation}

The two invariant masses are independent variables. Their kinematically
allowed region is determined by
\begin{equation}
    s_1\geq4m_{f'}^2,
    \qquad
    s_2\geq4m_{f'}^2,
\end{equation}
together with
\begin{equation}
    \sqrt{s_1}+\sqrt{s_2}\leq m_\phi.
\end{equation}
A convenient parametrisation of the integration region is therefore
\begin{equation}
    4m_{f'}^2
    \leq
    s_1
    \leq
    (m_\phi-2m_{f'})^2,
\end{equation}
and, for fixed $s_1$,
\begin{equation}
    4m_{f'}^2
    \leq
    s_2
    \leq
    (m_\phi-\sqrt{s_1})^2.
\end{equation}

\subsection{Factorised 4-body contribution}

Within the factorised treatment, the 4-body
contribution is
\begin{equation}
    \Gamma'_4
    =
    \frac{S}{2m_\phi}
    \int
    \sum_{\rm spins}|\mathcal M_4|^2
    d\Phi_4,
\end{equation}
with $S=1/4$.

Combining the matrix element and the phase-space measure gives
\begin{align}
    \Gamma'_4
    ={}&
    \frac{g'^2y'^4v_\phi^2}
         {4096\pi^5m_\phi^3}
    \int_{\mathcal R}ds_1\,ds_2\,
    \frac{
    s_1s_2\,
    \lambda^{1/2}(m_\phi^2,s_1,s_2)
    }{
    (s_1-M_{H'}^2)^2
    (s_2-M_{H'}^2)^2
    }
    \nonumber\\
    &\times
    \left(
    1-\frac{4m_{f'}^2}{s_1}
    \right)^{3/2}
    \left(
    1-\frac{4m_{f'}^2}{s_2}
    \right)^{3/2}.
\end{align}
The power $3/2$ arises from combining the massive spin sums with the
2-body phase-space factors,
\begin{equation}
    (s_i-4m_{f'}^2)
    \left(
    1-\frac{4m_{f'}^2}{s_i}
    \right)^{1/2}
    =
    s_i
    \left(
    1-\frac{4m_{f'}^2}{s_i}
    \right)^{3/2}.
\end{equation}

\subsection{Dimensionless form}

We define
\begin{equation}
    z\equiv\frac{m_\phi}{M_{H'}},
    \qquad
    \mu\equiv\frac{m_{f'}}{m_\phi},
    \qquad
    x_i\equiv\frac{s_i}{m_\phi^2}.
\end{equation}
For the benchmark relation
\begin{equation}
    m_{f'} = y'M_{H'} ,
\end{equation}
the variables $\mu$ and $y'$ are related by
\begin{equation}
    \mu=\frac{y'}{z},
    \qquad
    y'=z\mu.
\end{equation}

Kinematic accessibility requires
\begin{equation}
    m_\phi>4m_{f'},
\end{equation}
and therefore
\begin{equation}
    0<\mu<\frac14,
\end{equation}
or equivalently
\begin{equation}
    0<y'<\frac{z}{4}.
\end{equation}

In terms of the dimensionless variables, the decay width becomes
\begin{equation}
    \Gamma'_4
    =
    \frac{g'^2y'^4v_\phi^2}
         {4096\pi^5m_\phi}
    z^8J_4(z,\mu),
\end{equation}
where
\begin{align}
J_4(z,\mu)
={}&
\int_{4\mu^2}^{(1-2\mu)^2}dx_1
\int_{4\mu^2}^{(1-\sqrt{x_1})^2}dx_2
\,
\frac{
x_1x_2\,
\lambda^{1/2}(1,x_1,x_2)
}{
(1-z^2x_1)^2
(1-z^2x_2)^2
}
\nonumber\\
&\times
\left(
1-\frac{4\mu^2}{x_1}
\right)^{3/2}
\left(
1-\frac{4\mu^2}{x_2}
\right)^{3/2}.
\end{align}

Using
\begin{equation}
    \Gamma_{\phi\rightarrow HH}
    \simeq
    \frac{g^2v_\phi^2}{32\pi m_\phi},
\end{equation}
the 4-body reheating asymmetry is
\begin{equation}
    \epsilon_4(z,\mu)
    =
    \frac{1}{128\pi^4}
    \left(\frac{g'}{g}\right)^2
    z^{12}\mu^4J_4(z,\mu).
\end{equation}

For each fixed value of $z$, the largest double-off-shell
4-body reheating asymmetry is obtained by maximising over the
interval in which both dark Higgs propagators remain off shell.
Thus,
\begin{equation}
    \epsilon_4^{\rm max}(z)
    =
    \frac{1}{128\pi^4}
    \left(\frac{g'}{g}\right)^2
    z^{12}
    \max_{\mu\in{\cal I}(z)}
    \left[\mu^4 J_4(z,\mu)\right],
\end{equation}
with
\begin{equation}
    {\cal I}(z)=
    \begin{cases}
        \left(0,\frac14\right),
        & z<1,\\[2mm]
        \left(\dfrac{z-1}{2z},\frac14\right),
        & 1<z<2.
    \end{cases}
\end{equation}
The lower bound in the second line enforces
$m_\phi<M'_H+2m'_f$, so that the single-resonant
3-body configuration is kinematically inaccessible.

The endpoint $\mu=1/4$, or equivalently $y'=z/4$, corresponds to
$m_\phi=4m_{f'}$. At this point the 4-body phase space closes, and
the decay width vanishes. The physical maximum therefore occurs at an
intermediate value of the Yukawa coupling.

\subsection{Off-shell limit}

As a consistency check, consider the strongly hierarchical and
massless limit
\begin{equation}
    z\ll1,
    \qquad
    \mu\rightarrow0.
\end{equation}
The propagator factors reduce to
\begin{equation}
    (1-z^2x_i)^{-2}\longrightarrow1,
\end{equation}
while the massive phase-space factors approach unity. The remaining
dimensionless integral is
\begin{equation}
    J_4
    \longrightarrow
    \int_0^1dx_1
    \int_0^{(1-\sqrt{x_1})^2}dx_2\,
    x_1x_2\lambda^{1/2}(1,x_1,x_2).
\end{equation}
Evaluating the integral gives
\begin{equation}
    J_4
    \longrightarrow
    \frac{1}{720}.
\end{equation}
The decay width therefore reduces to
\begin{equation}
    \Gamma'_4
    \longrightarrow
    \frac{g'^2y'^4v_\phi^2}
         {2949120\,\pi^5m_\phi}
    \left(
    \frac{m_\phi}{M_{H'}}
    \right)^8 .
\end{equation}
This reproduces the expected strong power-law suppression of the
4-body channel in the deep off-shell regime, while the full
expression derived above remains valid away from the strongly
hierarchical limit and consistently incorporates the finite
dark fermion mass.

\bibliographystyle{JHEP} 
\bibliography{biblio}


\end{document}